\documentclass{aastex6}

\usepackage{amsfonts}
\usepackage[utf8x]{inputenc}
\usepackage{graphicx}	%load this before "epstopdf"
\usepackage{epstopdf}
\usepackage{bm}
\usepackage{amsmath}
\usepackage{natbib}	% use Natbib bibliography
\usepackage{xcolor}

\newcommand{\Rm}{\mathrm}

\begin{document}
	\title{Multi-fluid approach to high-frequency waves in plasmas: I. Small-amplitude regime in fully ionized medium}
	\shorttitle{High-frequency waves in plasmas}
	\shortauthors{Martínez-Gómez et al.}
	
	\author{David Martínez-Gómez\altaffilmark{1,2}, Roberto Soler\altaffilmark{1,2}, and Jaume Terradas\altaffilmark{1,2}}
	\altaffiltext{1}{Departament de Física, Universitat de les Illes Balears, 07122, Palma de Mallorca, Spain}
	\altaffiltext{2}{Institut d'Aplicacions Computacionals de Codi Comunitari (IAC3), Universitat de les Illes Balears, 07122, Palma de Mallorca, Spain}
	\email{david.martinez@uib.es}

\begin{abstract}
	Ideal MHD provides an accurate description of low-frequency Alfvén waves in fully ionized plasmas. However, higher frequency waves in many plasmas of the solar atmosphere cannot be correctly described by ideal MHD and a more accurate model is required. Here, we study the properties of small-amplitude incompressible perturbations in both the low and the high frequency ranges in plasmas composed of several ionized species. We use a multi-fluid approach and take into account the effects of collisions between ions and the inclusion of Hall's term in the induction equation. Through the analysis of the corresponding dispersion relations and numerical simulations we check that at high frequencies ions of different species are not as strongly coupled as in the low frequency limit. Hence, they cannot be treated as a single fluid. In addition, elastic collisions between the distinct ionized species are not negligible for high frequency waves since an appreciable damping is obtained. Furthermore, Coulomb collisions between ions remove the cyclotron resonances and the strict cut-off regions that are present when collisions are not taken into account. The implications of these results for the modelling of high-frequency waves in solar plasmas are discussed.
\end{abstract}	


\section{Introduction} \label{sec:intro}
	The pioneering work of \citet{1942Natur.150..405A} set the starting point of magnetohydrodynamics (MHD), which has become a remarkably successful theory for understanding the general properties of the most abundant state of matter in the universe, i.e., plasma. For instance, one of its fundamental predictions, the existence of magnetohydrodynamic waves driven by magnetic tension, was experimentally confirmed in a laboratory by \citet{1949Natur.164..145L} and corroborated later by \citet{1954PhRv...94..815L} and \citet{1959Natur.183.1652J}. Subsequent investigations have demonstrated the presence of this class of oscillations, now known as Alfvén waves, in Earth's aurora \citep{1988PhyS...38..841C}, in planetary ionospheres \citep{1960JGR....65.2233B,1981JGR....86..717G}, the solar wind \citep{1966PhysRevLett.17.207,1971JGR....76.3534B} or the interstellar medium \citep{1975ApJ...196L..77A,1996ApJ...465..775B}. \citet{2007Sci...317.1192T} and \citet{2007Sci...318.1574D} reported the detection of Alfvén waves in the solar atmosphere, although there is a debate about whether those observations are best interpreted in terms of Alfvén or kink waves \citep{2008ApJ...676L..73V,2009Sci...323.1582J,2012ApJ...753..111G}. In addition, they have been invoked as a mechanism for heating the solar corona \citep[e.g.,][]{1974SoPh...39..129W,2011Natur.475..477M} and may have a strong influence in many astrophysical phenomena like, e.g., the propagation of cosmic rays \citep{1969ApJ...156..445K} or the winds of hot stars \citep{1983ApJ...268L.127U}. Nevertheless, Alfvén waves, that correspond to the low range of frequencies, are only a subset of all the types of waves that can be found in plasmas \citep{1992wapl.book.....S}. 
	
	MHD has been used to investigate waves in a great variety of environments, like extragalactic jets \citep{1993A&A...279..351G}, protostellar winds \citep{1989MNRAS.236....1J},  molecular clouds \citep{1990ApJ...350..195P,2011MNRAS.415.1751M}, laboratory plasmas \citep{1952PhysRev.87.671,1999JGRA:JGRA14819} and fusion devices \citep{1971PlPh...13..258W,1974PhFl...17.1399C,2008PhPl...15e5501H}. However, the properties of high frequency waves, that in the solar atmosphere may be driven by small-scale magnetic activity in the chromospheric network and reconnection of field lines \citep{1992sws..coll....1A,1997SoPh..171..363T} or by cascading from low frequencies in the solar corona \citep{1983JGR....88.3923I,1987SoPh..109..149T}, are not accurately described by this theory and more general approaches are required. Among other simplifications, ideal MHD treats the plasma as if it were fully ionized, ignores the possible presence of neutral species, assumes that there is no resistivity, neglects Hall's term in the induction equation and considers all the ionized species together as a single fluid. Extensions to ideal MHD have addressed various of the ignored effects. For instance, the effect of partial ionization on waves in plasmas has been studied by, e.g., \citet{1956MNRAS.116..314P,2001ApJ...558..859D,2007A&A...461..731F,2010A&A...512A..28S} and \citet{2011A&A...534A..93Z}, although those authors did not consider each ionized species as a separate fluid. Investigations of fully ionized plasmas through the application of a multi-fluid description can be found in, e.g., \citet{1973Ap&SS..20..391W,1982JGR....87.5023I,2001paw..book.....C} and \citet{2002JGRA..107.1147H}, but in those works some issues have still been overlooked: in the former and the third, elastic collisions between the different ionized species are not taken into account, and the other two publications focus exclusively on the low frequency Alfvén waves. Still in the range of low frequencies, \citet{2007ApJ...661.1222L} and \citet{2008ApJ...682..667L} examined cases where the Wentzel–Kramers–Brillouin (WKB) approximation does not hold. The effect of collisions is included in the work of \citet{2010Rahbarnia}, although it studies the case of plasmas composed of only two distinct ionized species. In addition, waves and instabilities in anisotropic magnetized plasmas have been analyzed through 16-moment transport equations \citep{1979JPhD...12.1051D,1999JGR...104.9963O,2008A&A...489..769D}.
	
	Multi-fluid approaches to describe multicomponent plasmas have been commonly used in aeronomy and space physics \citep{1977RvGSP..15..429S,1982PlPh...24..389B}. For instance, such models have been extensively applied to the investigation of Earth's ionosphere \citep[see, e.g.,][]{1988AdSpR...8...69G,1989P&SS...37.1157K,1994JGR....99.2215D,1996RvGeo..34..311G} and the solar wind \citep{2004AdSpR..33..681O,2011SGeo...32....1E,2016SSRv..tmp...34A}. The study of the solar and stellar winds through the application of multi-fluid equations started with \citet{1966PhRvL..16..628S} and \citet{1968ApJ...151.1155H}, where separated heat equations for protons and electrons where considered. Later, \citet{1972SoPh...23..238L} included the proton temperature anisotropy and \citet{1973Ap&SS..20..401W} and \citet{2000A&A...359..983K} took into account the presence of several ions, but assuming isothermal ion flows, assumption that was then removed in \citet{2001A&A...369..222K}. These early models were one-dimensional (1D) and essentially hydrodynamic and did not explicitly involve Faraday's law. The inclusion of the induction equation is necessary in multi-dimensional models and it has been incorporated in works like \citet{2000JGR...10512675U,2001SoPh..199..371C,2003ApJ...598.1361V,2004JGRA..109.7102O,2004JGRA..109.7103L,2006JGRA..111.8106L,2014ApJ...782...81V} and \citep{2015MNRAS.454.3697M}. Moreover, taking into account Faraday's law allows the investigation of the angular momentum loss due to magnetized multi-ion stellar winds \citep{2006A&A...456..359L,2007ApJ...661..593L} and the effect of ion temperature anisotropy on that momentum loss \citep{2009A&A...494..361L}.
	
	When considering a multi-fluid plasma, the distinct species may interact with each other in two ways: through electro-magnetic fields, which only affect the charged particles, and by means of collisions. For low-frequency waves the friction due to collisions between the ionized species is usually ignored. The reason to do so is that the magnetic field produces a strong enough coupling between the different charged fluids so that they behave almost as a single fluid. Consequently, the friction force between ionized species can be neglected. This may not be true for high-frequency waves, at which each charged fluid may react to the perturbations of the electric and the magnetic fields in different time scales and the effects of the cyclotron motions should be taken into account. At the range of high frequencies the velocity drifts are not negligible and the frictional force may be of great relevance. This frictional dissipation of the energy carried by the waves may have an important role in the heating of the plasma.
	
	The present paper is the first of a series whose general goal is the investigation of waves in multi-component plasmas, with the ultimate objective of studying the heating that can arise from the complex interactions between the different species. To that aim we have developed a numerical code that implements a multi-fluid model that takes into account the elastic collisions between species and makes use of a generalized Ohm's law. Although the code is able to simulate the effects of partial ionization and nonlinearity, in this paper we focus on the simple case of fully ionized homogeneous plasmas and perturbations in the small-amplitude regime. Thus, here the heat transfer due to collisions, which has a strong dependence on the amplitudes of the velocity perturbations, is not expected to have a prominent role. Nevertheless, the momentum transfer between different ionized species may be of great influence in the properties of waves. The effects of partial ionization and nonlinearity are not addressed in the present initial article and will be explored in forthcoming papers of the series. 
	
	The inclusion of several ionized species, together with the consideration of Hall's term, produces the appearance of circularly polarized waves. We are interested in studying under which circumstances Coulomb collisions should be taken into account to properly describe the propagation of waves. In addition, we compare the multi-fluid approach with ideal MHD. The purpose is to check the applicability of MHD and to determine the conditions for which the multi-fluid description is necessary.

	The equations that describe the multi-fluid model are presented in Section \ref{sec:equations}. In Section \ref{sec:dispersion} we derive the general dispersion relation that characterize the properties of incompressible waves in a homogeneous medium and analyze the particular cases of two-ion and three-ion plasmas, with a focus in three distinct regions of the solar atmosphere: the upper chromospheric region, the lower corona and the solar wind at 1 astronomical unit (AU). In Section \ref{sec:simulations}, we show the results of simulations of the temporal evolution of small-amplitude perturbations and compare them with the predictions from the dispersion relations. Finally, the summary is presented in Section \ref{sec:concl}.

\section{Fluid equations for a multi-ion plasma} \label{sec:equations}
	The equations governing the mass, momentum and energy transport for species $s$ in a multi-component plasma read
	\begin{equation}\label{eq:cont_s}
		\frac{\partial n_{s}}{\partial t}+\nabla \cdot \left(n_{s}\bm{V_{s}}\right)=0,
	\end{equation}
	\begin{eqnarray}\label{eq:mom_s}
		\frac{\partial \left(m_{s}n_{s}\bm{V_{s}}\right)}{\partial t}&+&\nabla \cdot 
		\left(m_{s}n_{s}\bm{V_{s}}\bm{V_{s}}+P_{s}\mathbb{I}\right)=
		q_{s}n_{s}\left(\bm{E}+\bm{V_{s}}\times \bm{B}\right) \nonumber \\ &+&m_{s}n_{s}\bm{g}+\sum_{t \ne s}\bm{R_{s}^{st}},
	\end{eqnarray}
	\begin{equation}\label{eq:pres_s}
		\frac{\partial P_{s}}{\partial t}=-\left(\bm{V_{s}}\cdot \nabla \right)P_{s}
		-\gamma P_{s}\nabla \cdot \bm{V_{s}}+\left(\gamma -1\right)\sum_{t \ne s}Q_{s}^{st},
	\end{equation}
	\noindent where $m_{s}$ is the mass, $n_{s}$ is the number density, $\bm{V_{s}}$ is the velocity, $P_{s}$ is the gas pressure, $\mathbb{I}$ is the three-dimensional identity matrix, and $q_{s}=Z_{s}e$ is the electric charge (with $e$ the elementary charge and $Z_{s}$ the signed charge number). The acceleration of gravity is denoted by $\bm{g}$, the electric field by $\bm{E}$, the magnetic field by $\bm{B}$, and $\gamma$ is the adiabatic index.
	
	We note that, since this paper is devoted to the study of fully ionized plasmas, here neutrals are not included. The investigation of the effects of partial ionization will be carried out in a forthcoming paper. 
	
	The momentum transfer and the heat transfer due to elastic collisions between two species, $s$ and $t$, are defined as \citep[see][]{1977RvGSP..15..429S,1986MNRAS.220..133D}
	\begin{equation}\label{eq:rterms}
		\bm{R_{s}^{st}}\equiv \alpha_{st}\left(\bm{V_{t}}-\bm{V_{s}}\right)\Phi_{st}
	\end{equation}
	\textbf{and}
	\begin{equation}\label{eq:qterms}
		Q_{s}^{st}\equiv \frac{2\alpha_{st}}{\left(m_{s}+m_{t}\right)}\left[\frac{3}{2}k_{\Rm{B}}\left(T_{t}-T_{s}\right)\Psi_{st}+\frac{1}{2}m_{t}\left(\bm{V_{t}}-\bm{V_{s}}\right)^2\Phi_{st}\right],
	\end{equation}
	respectively, with $\alpha_{st}=\alpha_{ts}$ the friction coefficient of collisions between species $s$ and $t$, $k_{\Rm{B}}$ the Boltzmann constant and $T_{s}$ the temperature of species $s$. The functions $\Phi_{st}$ and $\Psi_{st}$ are correction factors that depend on the drift speed, $|\bm{V_{s}}-\bm{V_{t}}|$, and on the reduced thermal speed, $V_{therm} \equiv \sqrt{2k_{\Rm{B}}\left(m_{t}T_{s}+m_{s}T_{t}\right)/\left(m_{s}m_{t}\right)}$. Unless otherwise stated, along this work we use the values $\Phi_{st}=\Psi_{st}=1$, which is a good approximation when the drift speed is small compared to the reduced thermal speed \citep{1977RvGSP..15..429S}. The friction coefficient for collisions between ionized particles is given by \citep[see, e.g.,][]{1965RvPP....1..205B,Callen2006a}
	\begin{equation}
		\alpha_{st}=\frac{n_{s}n_{t}Z_{s}^2Z_{t}^2e^4\ln{\Lambda_{st}}}{6\pi\sqrt{2\pi}\epsilon_{0}^2m_{st}\left(k_{\Rm{B}}T_{s}/m_{s}+k_{\Rm{B}}T_{t}/m_{t}\right)^{3/2}},
	\end{equation} 
	where $\epsilon_{0}$ is the vacuum electrical permittivity and $m_{\Rm{st}} = m_{\Rm{s}}m_{t}/ \left(m_{s}+m_{t}\right)$ is the reduced mass. Coulomb's logarithm is defined as \citep[see, e.g.,][]{1962pfig.book.....S,2013A&A...554A..22V}
	\begin{equation}\label{eq:coulomb}
		\ln \Lambda_{st}=\ln\left[\frac{12\pi\epsilon_{0}^{3/2}k_{\Rm{B}}^{3/2}\left(T_{s}+T_{t}\right)}{|Z_{s}Z_{t}|e^3}\left(\frac{T_{s}T_{t}}{Z_{s}^2n_{s}T_{t}+Z_{t}^2n_{t}T_{s}}\right)^{1/2}\right].
	\end{equation} 
	 
	The system of Equations (\ref{eq:cont_s})-(\ref{eq:pres_s}), which ignores self-collisions and so does not take into account the effects of viscosity, must be completed with additional equations for the magnetic field and the electric field. The former is given by Faraday's law,
	\begin{equation}\label{eq:induction}
		\frac{\partial \bm{B}}{\partial t}=-\nabla \times \bm{E}.
	\end{equation}
	\noindent In turn, the electric field is obtained from the momentum equation of electrons \citep[see, e.g.,][]{2011A&A...529A..82Z,2014PhPl...21i2901K} (which are denoted by the index ``e'' and whose charge number is $Z_{\Rm{e}}=-1$), namely
	\begin{eqnarray}\label{eq:mom_e}
		\frac{\partial \left(\rho_{e}\bm{V_{e}}\right)}{\partial t}&+&\nabla \cdot 
		\left(\rho_{e}\bm{V_{e}}\bm{V_{e}}+P_{e}\mathbb{I}\right)=
		-en_{e}\left(\bm{E}+\bm{V_{e}}\times \bm{B}\right) \nonumber \\
		&+&\rho_{e}\bm{g}
		+\sum_{j \ne e}\bm{R_{e}^{ej}}.
	\end{eqnarray}

	If we consider that the variations of momentum of electrons are negligible (which is justified by their very small mass), then we can neglect the first two terms on the left-hand side of Equation (\ref{eq:mom_e}) and express the electric field as
	\begin{equation}\label{eq:mom_e2}
		\bm{E}=-\bm{V_{e}}\times \bm{B}-\frac{\nabla P_{e}}{en_{e}}+\frac{m_{e}}{e}\bm{g}+\frac{1}{en_{e}}\sum_{j \ne e}\bm{R_{e}^{ej}}.
	\end{equation}

	To ease later computations, we define a new velocity related to the ions, $\bm{V}$, as
	\begin{equation}\label{eq:vions}
		\bm{V}\equiv \frac{\sum_{i}^{M}{Z_{i}n_{i}\bm{V_{i}}}}{n_{e}},
	\end{equation}
	where $M$ is the number of ionized species and $n_{e}$ is given by the condition of quasi-neutrality, $\sum_{s}Z_{s}n_{s}\approx 0$, so $n_{e} \approx \sum_{i}^{M}Z_{i}n_{i}$. We note that this velocity is different from the center-of-mass velocity. 
	
	The current density is given by
	\begin{equation}\label{eq:current}
		\bm{j}=\sum_{s}q_{s}n_{s}\bm{V_{s}}, 
	\end{equation}
	but making use of Equation (\ref{eq:vions}) it can be expressed as
	\begin{equation}\label{eq:current_2}
		\bm{j}=en_{e}\left(\bm{V}-\bm{V_{e}}\right).
	\end{equation}

	Now we rewrite Equation (\ref{eq:current_2}) as $\bm{V_{e}}=\bm{V}-\bm{j}/en_{e}$ and insert this expression into Equation (\ref{eq:mom_e2}) to obtain the version of the generalized Ohm's law that is used in the numerical code, namely
	\begin{equation}\label{eq:ohm}
		\bm{E}=-\bm{V}\times \bm{B}+\frac{\bm{j}\times \bm{B}}{en_{e}}-\frac{\nabla P_{e}}{en_{e}}+\frac{m_{e}}{e}\bm{g}+\frac{1}{en_{e}}\sum_{j \ne e}\bm{R_{e}^{ej}}.
	\end{equation} 

	Hence, according to this multi-fluid approach the temporal evolution of a multi-ion plasma is described by a total of five equations for each species (continuity, three components of momemtum and pressure), usually known as 5-moment transport equations \citep[see, e.g.,][]{1977RvGSP..15..429S}, in addition to the pressure equation for electrons and the three components of Faraday's law. Due to the complexity of such a system, our general procedure will be to numerically compute the temporal evolution using the MolMHD code \citep{Bona20092266}, which originally solved the ideal MHD equations but has been extended by adding a new module that incorporates the multi-fluid equations given above. However, for some specific cases we can obtain analytical expressions for the dispersion relations that characterize the propagation of linear waves in the plasma and we can compare the results provided by those two different approaches.
	
	This model allows for the examination of effects that cannot be addressed by the single-fluid approach, but it also has some limitations. By treating each species as a fluid and resorting to the 5-moment transport equations approximation, the velocity distribution functions (VDFs) of the species are assumed to be Maxwellian \citep{1977RvGSP..15..429S}. But at frequencies comparable to the ions gyrofrequencies this assumption may not hold due to the wave-particle interactions. Such interactions are better described by kinetic models \citep{2002JGRA..107.1147H} and there are features, like the secondary proton beam in the fast solar wind, that cannot be studied with the fluid approach. However, kinetic methods are much more complicated and computationally costly and the departure of the VDFs from a Maxwellian may be only of significance when considering large-amplitude waves. Another limitation is that the 5-moment approximation neglects the effects of anisotropic pressures, thermal diffusion and thermal conduction. Such shortcoming can be resolved by considering stress and heat flow in higher-order moment approximations, but at the expense of more complex calculations. For the time being, we focus in plasmas with isotropic temperatures and we do not take into account the heat flow.

\section{Dispersion relation of linear waves} \label{sec:dispersion}
	As the target of this section is to describe the behavior of incompressible perturbations in a homogeneous plasma, we can ignore the various continuity and pressure equations and retain only the equations governing momentum, and the magnetic and electric fields. Moreover, the battery term in Ohm's law, i.e., the term related to the gradient of the electronic pressure, is expected to have little influence under the chosen conditions for the plasma, so we can drop it. And since we do not take into account collisions between the ions and electrons,  resistivity can also be disregarded, i.e., the last term on the right-hand side of Equation (\ref{eq:ohm}) can be ignored. In addition, we consider wavelengths that are much shorter than the scale height due to gravity, so we can also neglect this effect as well. Hence, Ohm's law can be now written as
	\begin{equation}\label{eq:ohm_2}
		\bm{E}=-\bm{V}\times \bm{B}+\frac{\bm{j}\times \bm{B}}{en_{e}}.
	\end{equation}

	To obtain a dispersion relation for the waves we linearize Equations (\ref{eq:mom_s}),(\ref{eq:induction}) and (\ref{eq:ohm_2}) assuming that each of the variables can be separated into two terms: a constant equilibrium value, denoted by the subscript ``0'', plus a small-amplitude perturbation, denoted by the subscript ``1''. Hence,
	\begin{equation} \label{velocity_s}
		\bm{V_{s}}=\bm{V_{s,0}}+\bm{V_{s,1}}, \ \bm{E}=\bm{E_{0}}+\bm{E_{1}}, \ \bm{B}=\bm{B_{0}}+\bm{B_{1}}. 
	\end{equation}

	We consider a uniform static background so the temporal and spatial derivatives of the equilibrium values are equal to zero and $\bm{V_{s,0}}=0$. Then, $\bm{V_{s}}=\bm{V_{s,1}}$ and we can drop the subscript ``1'' from the velocity variables from here on. In addition, from Ampère's law the current density can be expressed as a function of the magnetic field perturbation only, namely
	\begin{equation}\label{eq:ampere}
		\bm{j}=\frac{\nabla \times \bm{B}}{\mu_{0}}=\frac{\nabla \times \left(\bm{B_{0}}+\bm{B_{1}}\right)}{\mu_{0}}=\frac{\nabla \times 	\bm{B_{1}}}{\mu_{0}}.
	\end{equation}
	\noindent with $\mu_{\Rm{0}}$ the magnetic permeability. The linearization process, in which we neglect second order products of the perturbed quantities and retain only those of first order, yields the following equations:
	\begin{equation}\label{eq:ohm_3}
		\bm{E_{1}}=-\bm{V}\times \bm{B_{0}}+\frac{\left(\nabla \times \bm{B_{1}}\right)\times \bm{B_{0}}}{en_{e}\mu_{0}},
	\end{equation}
	\begin{eqnarray} \label{eq:mom_s_2}
		\frac{\partial \bm{V_{s}}}{\partial t}&=&\frac{Z_{s}e}{m_{s}}\left(\bm{E_{1}}+\bm{V_{s}}\times \bm{B_{0}}\right) \nonumber \\
		&+&\sum_{t\ne s}\nu_{st}\left(\bm{V_{t}}-\bm{V_{s}}\right) \ \ \ s,t \in \left(1,\cdots,M\right),
	\end{eqnarray}
	\begin{equation}\label{eq:induction_2}
		\frac{\partial \bm{B_{1}}}{\partial t}=-\nabla \times \bm{E_{1}},
	\end{equation}
	\noindent where $\nu_{st}=\alpha_{st}/\rho_{s}$ is the collision frequency between two species $s$ and $t$, and $\rho_{s}=m_{s}n_{s}$ is the mass density of species $s$.

	Then we insert Equation (\ref{eq:ohm_3}) into Equations (\ref{eq:mom_s_2}) and (\ref{eq:induction_2}), and perform a normal mode analysis expressing the perturbations as proportional to $\exp \left(-i\omega t\right)$, where $\omega$ is the frequency. Moreover, we perform a Fourier analysis in space. Hence, we impose that the perturbations are also proportional to $\exp \left(i\bm{k} \cdot \bm{r}\right)$, where $\bm{k}$ is the wave vector and $\bm{r}$ the position vector. For simplicity, we choose a reference frame in which $\bm{B_{0}}=\left(B_{x},0,0\right)$, $\bm{k}=\left(k_{x},0,0\right)$ and assume that the motions of the perturbations are in the $y$ and $z$ directions. We arrive at a set of twelve equations where the $y$ and $z$ components are not independent from each other but they are coupled and, in principle, it is not possible to study the behavior of the perturbations in one direction exclusively while ignoring the other direction. We could now rearrange the different terms in those equations to express the system in a matrix form and get the dispersion relation by equating to zero the determinant of the coefficient matrix. However, we find it more convenient to perform some additional steps that, in the end, simplify the calculations and give a better insight on the results. Thus, as in \citet{1992wapl.book.....S} or \citet{2001paw..book.....C}, we define the following variables  
	\begin{equation} \label{eq:polarized}
		V_{s,\pm}=V_{s,y}\pm i V_{s,z}, \ B_{1,\pm}=B_{1,y}\pm i B_{1,z},
	\end{equation}
	\noindent that correspond to the left-hand (+) and right-hand (-) circular polarization of the original variables. Such definitions allow us to get to a new set of equations:
	\begin{eqnarray} \label{eq:vs+-}
		\omega V_{s,\pm}&=&\Omega_{s}\left[\pm \left(V_{s,\pm}-V_{\pm}\right)- \frac{k_{x}}{e n_{e}\mu_{0}}B_{1,\pm}\right] \nonumber \\
		&+&i \sum_{t \ne s}\nu_{st}\left(V_{t,\pm}-V_{s,\pm}\right) \ \ \ s,t \in \left(1,\cdots,M\right),
	\end{eqnarray}
	\begin{equation} \label{eq:magnetic+-}
		\omega B_{1,\pm}=-k_{x}B_{x}V_{\pm} \mp \frac{k_{x}^2 B_{x}}{e n_{e} \mu_{0}}B_{1,\pm},
	\end{equation}
	\noindent where $\Omega_{s}=Z_{s}eB_{x}/m_{s}$ is the cyclotron frequency of the ion $s$. Thanks to those manipulations we have transformed a problem of eight coupled equations into two independent systems of four coupled equations. Now, we can rearrange the terms in the above formulas and write both problems in matrix form:
	\begin{equation} \label{eq:matricial}
		A_{\pm}\cdot \bm{u_{\pm}}=0,
	\end{equation}
	\noindent with $\bm{u_{\pm}}=\left(V_{1,\pm},\cdots,V_{\Rm{M},\pm},B_{1,\pm}\right)$. The coefficient matrices $A_{\pm}$ for the case with $M=3$ are shown in Appendix \ref{app:A}. The dispersion relation for the circularly polarized incompressible waves is given by
	\begin{equation} \label{eq:dispersion}
		\mathcal{D}_{\pm}(\omega,k_{x}) \equiv \det A_{\pm}=0.
	\end{equation}

	The dispersion relation contains a great number of parameters (e.g., several collision frequencies, the number densities of each species, the cyclotron frequencies, the wavenumber of the perturbation, etc.). Thus, it is not easy to discriminate the effect of each one of those parameters on the final results. For this reason, it is convenient to begin with a simpler situation that allow us to focus on some of those effects before dealing with the general case. In Section \ref{sec:dispersion_two-ion} we consider a plasma made of two different ionized species, which considerably reduces the number of parameters involved in the analysis. The more general study of waves in a three-ion plasma is left for Section \ref{sec:dispersion_three-ion}.

\subsection{Waves in a two-ion plasma} \label{sec:dispersion_two-ion}
	To obtain the dispersion relation for incompressible waves in a two-ion plasma we set $M=2$. Then, we directly combine Equations (\ref{eq:vs+-}) and (\ref{eq:magnetic+-}) or solve the characteristic equation of the 3$\times$3 matrix that arises from that system. The outcome is
	\begin{eqnarray} \label{eq:dr_2ions_a}
		&\omega_{\pm}^2&\big[Z_{2}n_{2}\left(\rho_{1}\rho_{2}\left(\omega_{\pm} \mp \Omega_{1}\right)+i \alpha_{12}\left(\rho_{1}+\rho_{2}\right)\right) \nonumber \\ &+&Z_{1}n_{1}\big(\rho_{1}\rho_{2}\left(\omega_{\pm} \mp \Omega_{2}\right)+i\alpha_{12}\left(\rho_{1}+\rho_{2}\right)\big)\big] \nonumber \\ 
		&\pm& \frac{B_{x}k_{x}^2}{e \mu_{0}}\big[\rho_{1}\rho_{2}\left(\omega_{\pm}\mp \Omega_{1}\right)\left(\omega_{\pm} \mp \Omega_{2}\right) \nonumber \\
		&+&i \alpha_{12}\left(\rho_{1}\left(\omega_{\pm} \mp \Omega_{1}\right)+\rho_{2}\left(\omega_{\pm} \mp \Omega_{2}\right)\right)\big]=0,
	\end{eqnarray}
	which becomes
	\begin{eqnarray} \label{eq:dr_2ions_b}
		&\omega_{\pm}^2&\left[Z_{2}n_{2}\left(\omega_{\pm} \mp \Omega_{1}\right)+Z_{1}n_{1}\left(\omega_{\pm} \mp \Omega_{2}\right)\right] \nonumber \\ &\pm& \frac{B_{x}k_{x}^2}{e \mu_{0}}\left(\omega_{\pm}\mp \Omega_{1}\right)\left(\omega_{\pm}\mp \Omega_{2}\right)=0
	\end{eqnarray}
	when the effect of elastic collisions between the two ions is not included \citep{1973Ap&SS..20..391W,2001paw..book.....C}. An important goal of the present study is to determine whether collisions between different ions have a relevant role. Therefore, we shall compare the results of Equations (\ref{eq:dr_2ions_a}) and (\ref{eq:dr_2ions_b}). These dispersion relations allow us to study waves excited both by an impulsive driver or by a periodic driver, depending on whether we solve them as functions of the wavenumber, $k_{x}$ or as functions of the frequency, $\omega_{\pm}$, respectively. Contrary to the predictions of ideal MHD, where the results for the impulsive driver are equivalent to those for the periodic one, here the two distinct classes of waves manifest different properties. Equations (\ref{eq:dr_2ions_a}) and (\ref{eq:dr_2ions_b}) are third-order polynomials in $\omega_{\pm}$ but second-order polynomials in $k_{x}$. Consequently, three solutions can be obtained for waves generated by an impulsive driver and only two solutions appear when the driver is periodic, while ideal MHD gives two solutions for both kinds of drivers.
	
	An investigation on the effect of collisions on waves in a two-ion plasma has already been carried out by \citet{2010Rahbarnia}, although those authors focused only on waves generated by a periodic driver. In the present paper we analyze the two kinds of drivers and explore some results of the case with a periodic driver that have not been shown in \citet{2010Rahbarnia}, namely the amplitude ratios of the velocities of the ions and the quality factor of the perturbations.

\subsubsection{Impulsive driver} \label{sec:impulsive_2ions}
	The solutions to the dispersion relations may be complex. To study waves generated by an impulsive driver, we assume that $k_{x}$ is real and let $\omega_{\pm}$ be complex. Thus, each one of the three possible solutions can be written as $\omega=\omega_{R}+i \omega_{I}$, where $\omega_{R}$ is the actual frequency of oscillation and $\omega_{I}$ represents the damping rate of the perturbations when its value is negative or the growth rate when it is positive. In this work we expect that $\omega_{I} \le 0$ always, since there is no physical process in our model that could produce an instability.

	In general, we solve Equations (\ref{eq:dr_2ions_a}) and (\ref{eq:dr_2ions_b}) numerically, but there are certain limits where we can obtain some analytical expressions. For instance, when $k_{x} \to 0$, from Equation (\ref{eq:dr_2ions_a}) we find that one of the three possible solutions is
	\begin{equation} \label{eq:sol_cyclotron}
		\omega_{\pm}\approx \pm \widetilde{\Omega}-i \alpha_{12}\frac{\rho_{1}+\rho_{2}}{\rho_{1}\rho_{2}},
	\end{equation}
	where 
	\begin{equation} \label{eq:weighted_omega}
		\widetilde{\Omega}=\frac{Z_{2}n_{2} \Omega_{1}+Z_{1}n_{1}\Omega_{2}} {Z_{1}n_{1}+Z_{2}n_{2}}
	\end{equation}
	is the weighted average cyclotron frequency. This is a mode that appears in the multi-fluid description but is absent in ideal MHD. To find the other two solutions of the dispersion relation it is more convenient to perform some algebraic manipulations and rewrite Equation (\ref{eq:dr_2ions_a}) as
	\begin{equation} \label{eq:dr_2ions_c}
		\omega_{\pm}^{2}\pm \omega_{\Rm{A}}^2 \frac{\left[\left(\frac{\omega_{\pm}}{\Omega_{1}}\mp 1\right)\left(\frac{\omega_{\pm}}{\Omega_{2}} \mp 1\right)+i\Gamma\right]}{\left(\frac{\omega_{\pm}}{\widetilde{\Omega}} \mp 1+ i\frac{\alpha_{12}}{\widetilde{\Omega}} \frac{\rho_{1}+\rho_{2}}{\rho_{1}\rho_{2}}\right)}=0,
	\end{equation} 
	where $\omega_{\Rm{A}}=k_{x}c_{\Rm{A}}$ is the Alfvén frequency, $c_{\Rm{A}}=B_{x}/\sqrt{\mu_{0}\left(\rho_{1}+\rho_{2}\right)}$ is the two-ion Alfvén speed and $\Gamma$ is proportional to the friction coefficient (the full expression of $\Gamma$ can be found in the appendices, Equation (\ref{eq:Gamma})).

	When the elastic collisions between the ions are not considered, Equation (\ref{eq:dr_2ions_c}) reduces to
	\begin{equation} \label{eq:dr_2ions_d}
		\omega_{\pm}^{2}\pm \omega_{\Rm{A}}^2 \frac{\left[\left(\frac{\omega_{\pm}}{\Omega_{1}}\mp 1\right)\left(\frac{\omega_{\pm}}{\Omega_{2}} \mp 1\right)\right]}{\left(\frac{\omega_{\pm}}{\widetilde{\Omega}} \mp 1\right)}=0,
	\end{equation}
	that is equivalent to Equation (\ref{eq:dr_2ions_b}) and from which it is easy to find the remaining two solutions for the low wavenumber limit: if we assume that $\omega$ is much smaller than the cyclotron frequencies ($\Omega_{1}$, $\Omega_{2}$, and $\widetilde{\Omega}$), we obtain two modes with $\omega \approx \pm \omega_{\Rm{A}}$ for each polarization. These are the classic Alfvén waves. Hence, as expected, the multi-fluid description consistently recovers the classic Alfvén waves found in ideal MHD when the wave frequency is low and the collisions are neglected.

	Another limit that we can recover from Equation (\ref{eq:dr_2ions_a}) is that of Alfvén waves in a single-ion plasma. Setting one of the number densities to be equal to zero we arrive at
	\begin{equation} \label{eq:dr_1ion}
		\omega_{\pm}^2 \pm \frac{\omega_{\Rm{A}s}^2}{\Omega_{s}}\omega_{\pm}-\omega_{\Rm{A}s}^2=0,
	\end{equation}
	where $\omega_{\Rm{A}s}=k_{x}c_{\Rm{A}s}$ and $c_{\Rm{A}s}=B_{x}/ \sqrt{\mu_{0}\rho_{s}}$ are the Alfvén frequency and speed for a single-ion plasma, respectively. This is the dispersion relation for Alfvén waves in Hall MHD \citep{1960RSPTA.252..397L,2001paw..book.....C}. The corresponding ideal MHD formula is obtained when $\omega_{\pm} \ll \Omega_{s}$ and the second term of the equation can be ignored.

	After the validity of the dispersion relation has successfully been checked in various known limits, now we return to the general scenario and study the dependence of Equations (\ref{eq:dr_2ions_a}) and (\ref{eq:dr_2ions_b}) on any arbitrary wavenumber. These formulas are expressed in a way that can be applied to a great variety of plasmas: the ions in the fluid can be either positive or negative and either heavy or light. Nonetheless, we put our attention on plasmas that can be found in the solar atmosphere. More precisely, we focus on the following three cases: the upper chromospheric region, the  lower solar corona, and the solar wind at 1 AU. Although those three plasmas share the feature that they are mostly composed of hydrogen and helium, their physical conditions (temperatures, densities and magnetic fields) are quite different and it is interesting to compare the results obtained for each environment. In addition, we compare those results with the predictions given by the ideal MHD theory.
	\begin{figure}
				\centering
				\includegraphics{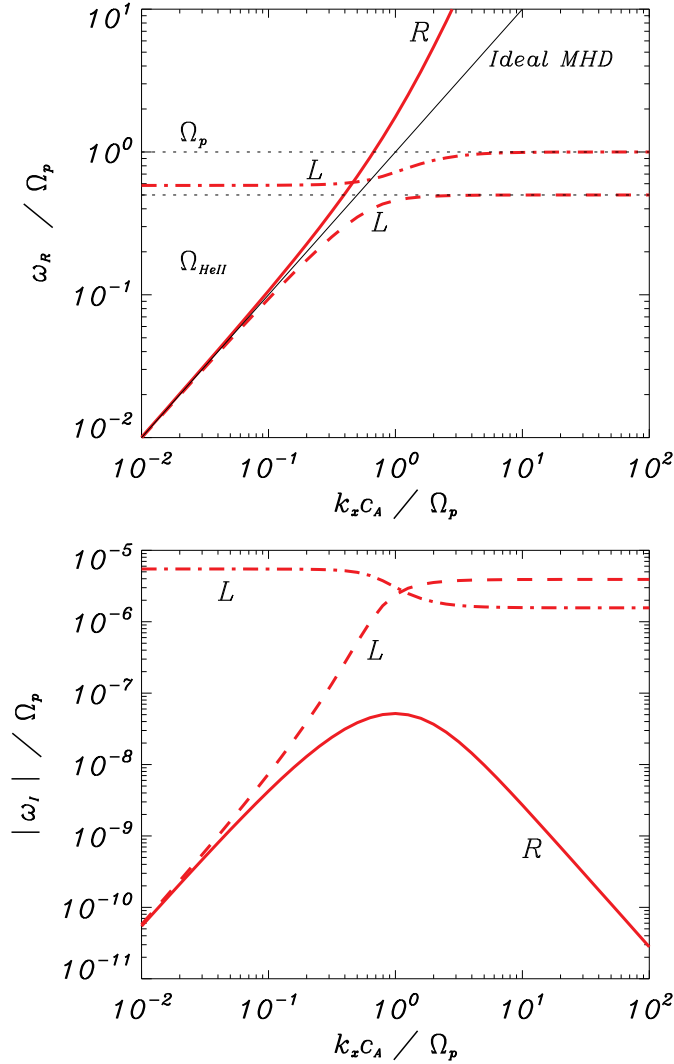}
%		\plotone{2ions_corona.eps}
		\caption{Solutions of the dispersion relations for a two-ion plasma with coronal conditions: $n_{p}=2.5 \times 10^{14} \ \Rm{m}^{-3}$, $n_{p\Rm{He}\textsc{iii}}=0.1n_{p}$, $B_{x}=10 \ \Rm{G}$, $T_{p}=T_{\Rm{He}\textsc{iii}}=10^6 \ \Rm{K}$, and $\nu_{p\Rm{He}\textsc{iii}}=0.15 \ \Rm{Hz}$. Top: normalized frequency, $\omega_{R}/\Omega_{p}$, as a function of the normalized wavenumber, $k_{x}c_{\Rm{A}}/\Omega_{p}$. Bottom: absolute value of the normalized damping, $|\omega_{I}|/\Omega_{p}$ as a function of $k_{x}c_{\Rm{A}}/\Omega_{p}$. The red lines correspond to the solutions from Equation (\ref{eq:dr_2ions_a}) (solid line: R mode; dashed and dot-dashed lines: L modes) and the black thin line represents the ideal MHD results ($\omega=\omega_{\Rm{A}}$). The dotted lines on the top panel represent the cyclotron frequency of each ion, with $\Omega_{p}>\Omega_{\Rm{He}\textsc{iii}}$.}
		\label{fig:2ions_wR_wI}
	\end{figure}

	We start by analyzing a plasma with coronal conditions, where the main components are protons, denoted by the subscript ``$p$'', and doubly ionized helium, ``$\Rm{He} \ \textsc{iii}$''.
	Some typical parameters for the lower solar corona are $n_{p}=2.5 \times 10^{14} \ \Rm{m}^{-3}$, $n_{\Rm{He}\textsc{iii}}=0.1n_{p}$, $B_{x}=10 \ \Rm{G}$, and $T=10^6 \ \Rm{K}$ \cite[see, e.g.,][]{1999JGR...104.9709F,1999ApJ...523..812S,2005A&A...435.1123W}. With these conditions the Alfvén speed is $c_{\Rm{A}}\approx 1160 \ \Rm{km}s^{-1}$, the collision frequencies are $\nu_{p\Rm{He}\textsc{iii}}=0.15 \ \Rm{Hz}$ and $\nu_{\Rm{He}\textsc{iii}p}=0.39 \ \Rm{Hz}$. The cyclotron frequencies are $\Omega_{p}\approx 96000 \ \Rm{rad \ s^{-1}}$, $\Omega_{\Rm{He}\textsc{iii}}\approx 48000 \ \Rm{rad \ s^{-1}}$, and $\widetilde{\Omega}=55878.1 \ \Rm{rad \ s^{-1}}$; hence $\nu_{st}\ll \Omega_{s}/(2\pi)$. The results of this study are shown on Figure \ref{fig:2ions_wR_wI}: the normalized real part of the frequency, $\omega_{R}/\Omega_{p}$, and the absolute value of the normalized damping, $|\omega_{I}|/\Omega_{p}$, as functions of the normalized wavenumber, $k_{x}c_{\Rm{A}}/\Omega_{p}$, are displayed on the top and the bottom panels, respectively.
	
	Only the modes with $\omega_{R}>0$ are depicted: two of them correspond to the left-hand polarization (and are denoted by the letter L) while the third belongs to the right-hand polarization (R). The region with $\omega_{R}<0$ would be symmetric with respect to the horizontal axis but the polarization of the modes would be exchanged. In addition, for the sake of clarity, only the solutions from the dispersion relation that takes collisions into account, Equation (\ref{eq:dr_2ions_a}), are shown. The reasons to do so are that no clear differences between the results from the two dispersion relations are appreciable in the real part of the frequency, i.e., collisions have a small effect on the frequency of oscillation of the waves, and that $\omega_{I}=0$ when friction is not considered. 

	Inspecting the top panel of Figure \ref{fig:2ions_wR_wI} we find that when $k_{x}c_{\Rm{A}}/\Omega_{p}\ll 1$ the multi-fluid result presents two branches with clearly distinct behavior: one of this branches (the red dot-dashed line) corresponds to the L mode associated with the cyclotron frequencies of the ions and whose frequency is given by Equation (\ref{eq:sol_cyclotron}); note that there is no ideal MHD solution related to this high-frequency mode, so it appears exclusively when several fluids are considered.
	
	In the second branch, we find the remaining L mode (dashed line) and the only R mode with $\omega_{R}>0$ (red solid line). These modes have almost the same value of the frequency, $\omega_{R}\approx \omega_{\Rm{A}}$, i.e., the frequency of oscillation is independent of the direction of polarization at this limit. At higher wavenumbers, this second branch splits into two and the corresponding waves become dispersive: their phase speeds are not independent of the wavenumber, contrary to low frequency Alfvén waves, that are non-dispersive. The roots of Equation (\ref{eq:dr_2ions_a}) start diverging from the ideal solution when $k_{x}c_{\Rm{A}}/\Omega_{p}\gtrsim 0.1$: the L mode, with $\omega_{R}<\omega_{\Rm{A}}$, tends to the cyclotron frequency of the more massive ion, $\Omega_{\Rm{He}\textsc{iii}}$; the R mode, with $\omega_{R}>\omega_{\Rm{A}}$, keeps increasing its frequency without converging to any limit in the range of wavenumbers explored in this study. Due to their behavior when $k_{x}c_{\Rm{A}}/\Omega_{p}\gtrsim 1$, the two L modes are commonly known as ion cyclotron waves and the R mode is known as the ion whistler wave \citep[see, e.g.,][]{2001paw..book.....C}.

	On the bottom panel of Figure \ref{fig:2ions_wR_wI} we see that collisions cause a clearly different damping on each mode of oscillation. The left-handed high-frequency modes are the most affected, while the damping on the Alfvénic modes and the high-frequency whistler wave is almost negligible. This variety of behaviors is caused by the velocity amplitude ratios and the phase shifts associated to each mode, issues that are analyzed later in this section.      

	Some useful parameters to analyze the behavior of waves are the quality factor, denoted by $Q \equiv 1/2|\omega_{R}/\omega_{I}|$ and the damping time, defined as $\tau \equiv 1/|\omega_{I}|$. The quality factor gives a measure of the relevance of the damping. If $Q>1/2$, the perturbation is said to be underdamped: it oscillates but its amplitude decreases with time; in the limit case of $Q \to \infty$ there is no damping at all. If $Q \le 1/2$ the wave is overdamped (with the special situation of $Q=1/2$ known as critically damped): the damping dominates the behavior of the perturbation. When $Q=0$ the mode is evanescent: there is no oscillation and the amplitude of the perturbation decays exponentially with time. In turn, the damping time represents the time interval in which the amplitude is reduced by a factor $1/e\approx 0.368$.
	
	From the results shown in Figure \ref{fig:2ions_wR_wI} it can be checked that for all modes $Q\gg 1$, i.e., the pertubations are extremely underdamped and the effect of collisions is almost irrelevant during a single period. However, for longer times the damping may not be negligible. For instance, if we set $k_{x}=\pi/10^5 \ \Rm{m}^{-1}$, we obtain the following damping times:  $\tau_{\Rm{-}}\approx \tau_{\Rm{+,1}}\approx 1.28 \times 10^{8} \ \Rm{s}$ and $\tau_{\Rm{+,2}}\approx 1.9 \ \Rm{s}$ (we remind the reader that the subscript ``+'' refers to the left-hand modes and ``-'' to the right-hand ones). This means that after a few seconds the perturbation related to the latter mode will have vanished and only the other two modes of oscillation will remain, almost undamped as if there were no collisions. For much higher wavenumbers the situation is different. For instance, if $k_{x}=\pi/10 \ \Rm{m}^{-1}$ (which corresponds to the normalized value $k_{x}c_{\Rm{A}}/\Omega_{p}\approx 3.8$), the damping times are $\tau_{-}\approx 670 \ \Rm{s}$, $\tau_{\Rm{+,1}}\approx 2.1 \ \Rm{s}$, and $\tau_{\Rm{+,2}}\approx 6.34 \ \Rm{s}$. Thus, the perturbations associated with the ion cyclotron modes will disappear after a few tens of seconds, while the whistler wave will have a considerably longer lifespan. If collisions between the two ionized species are not taken into account, none of the modes attenuates with time. Although the present linear analysis cannot capture this effect, the wave energy dissipated during the damping is expected to be deposited in the plasma and so to contribute to its heating.

	The next step in our work is to repeat the previous study but for a plasma with solar wind conditions. The solar wind at 1 AU can be described by the following set of parameters \citep[see, e.g.,][]{2001GeoRL..28.2767A,2003ApJ...591.1257L,2004prma.book.....G}: $n_{p}=10^7 \ \Rm{m}^{-3}$, $n_{\Rm{He}\textsc{iii}}=5 \times 10^5 \ \Rm{m}^{-3}$, $B_{x}\approx 5 \times 10^{-5} \ \Rm{G}$ and $T\approx 10^5 \ \Rm{K}$, which yields an Alfvén speed of $c_{\Rm{A}} \approx 31 \ \Rm{km \ s^{-1}}$, collision frequencies on the order of $10^{-7} \ \Rm{Hz}$ and cyclotron frequencies given by $\Omega_{p}\approx 0.479 \ \Rm{rad} \ \Rm{s}^{-1}$ and $\Omega_{\Rm{He}\textsc{iii}} \approx 0.239 \ \Rm{rad} \ \Rm{s}^{-1}$. Then, the results provided by the dispersion relations are qualitatively identical to those portrayed in Figure \ref{fig:2ions_wR_wI} for the solar corona, with the difference that now the collision frequencies are so minute that the solar wind can be considered a completely collisionless fluid from the perspective of this work. To support this statement we choose a wavenumber $k_{x}=\pi/10^5 \ \Rm{m}^{-1}$ (that corresponds to the normalized wavenumber $k_{x}c_{\Rm{A}}/\Omega_{p} \approx 2.1$) and check that all the damping times predicted by the dispersion relations are on the order of $10^6 \ \Rm{s}$ or larger. For lower values of $k_{x}c_{\Rm{A}}/\Omega_{p}$, the damping times are even greater. 
	
	We note that the approximation used for the functions $\Phi_{st}$ and $\Psi_{st}$ in Equations (\ref{eq:rterms}) and (\ref{eq:qterms}) is not strictly valid in this environment. In the solar wind at 1 AU the drift speed may be comparable to the reduced thermal speed and, hence, it would be more appropriate to employ the more general expressions for $\Phi_{st}$ and $\Psi_{st}$ given by \citet{1977RvGSP..15..429S}. However, due to the extremely low value of the collision frequencies, the application of those more realistic formulas would not modify in a remarkable way the results explained in the lines above.
	
	Nevertheless, it is still interesting to compare the results of the multi-fluid approach with those from ideal MHD. From Figure \ref{fig:2ions_wR_wI} we know that the ideal MHD description is not accurate enough when $k_{x}c_{\Rm{A}}/\Omega_{p}\gtrsim 0.1$. From that expression we can calculate what is the minimum wavelength for a perturbation to be reasonably well described by ideal MHD. We remind that $k_{x}=2\pi/\lambda$, where $\lambda$ is the wavelength. Then, we get
	\begin{equation} \label{eq:lambda_min}
		\lambda > \lambda_{c} \equiv \frac{2\pi c_{\Rm{A}}}{0.1 \Omega_{p}}.
	\end{equation}
	The solar wind parameters give a critical wavelength of $\lambda_{c} \approx 4 \times 10^{3} \ \Rm{km}$. The respective value for the solar corona is $\lambda_{c} \approx 750 \ \Rm{m}$.

	Next, we apply our model to the upper chromospheric region. We use the parameters given by the model F of \citet{1993ApJ...406..319F} for a height of $2000 \ \Rm{km}$ over the top of the photosphere: $n_{p}=10^{17} \ \Rm{m}^{-3}$, $n_{\Rm{He}\textsc{ii}}=10^{16} \ \Rm{m}^{-3}$, and $T=10^4 \ \Rm{K}$. A typical value of the magnetic field at that height is $B_{x}=35 \ \Rm{G}$. Note that in this case the second ionized species is not the doubly ionized helium but singly ionized helium, $\Rm{He} \ \textsc{ii}$, because the temperature is not large enough for helium to be fully ionized. The Alfvén speed is $c_{\Rm{A}}\approx 204 \ \Rm{km \ s^{-1}}$, the cyclotron frequencies $\Omega_{p}=335268 \ \Rm{rad \ s^{-1}}$, $\Omega_{\Rm{He}\textsc{ii}}=83817.1 \ \Rm{rad \ s^{-1}}$, and $\widetilde{\Omega}=106676 \ \Rm{rad \ s^{-1}}$, and the collision frequencies $\nu_{\Rm{pHe}\textsc{ii}} \approx 8500 \ \Rm{Hz}$ and $\nu_{\Rm{He}\textsc{ii}p} \approx 21300 \ \Rm{Hz}$.
	
	Again, Figure \ref{fig:2ions_wR_wI} can be used to describe the qualitative properties of waves in this plasma, but we need to bear in mind that now the collision frequencies are much larger than in the previous scenarios: now $\nu_{st}\simeq \Omega_{s}/(2\pi)$ and the damping produced by this interaction is considerably greater than before. In a plasma with coronal conditions, the maximum normalized damping we have found is $|\omega_{I}|/\Omega_{p}\approx 2 \times 10^{-5}$; if we apply the chromospheric conditions we get that such maximum value is given by $|\omega_{I}|/\Omega_{p}\approx 0.1$. In the same manner as in the previous cases, we first compute the solutions to Equation (\ref{eq:dr_2ions_a}) for a wavenumber of $k_{x}=\pi/10^5 \ \Rm{m^{-1}}$. Thus, the Alfvén frequency is $\omega_{\Rm{A}}=6.41 \ \Rm{rad} \ \Rm{s}^{-1}$ (so $k_{x}c_{\Rm{A}}/\Omega_{p} \approx 2 \times 10^{-5}$) and the dispersion relation yields $\omega_{-}=\omega_{+,1}\approx 6.41-i9.27\times 10^{-6} \ \Rm{rad} \ \Rm{s}^{-1}$ and $\omega_{+,2}\approx 106676-i29750 \ \Rm{rad} \ \Rm{s}^{-1}$. The quality factor for the Alfvénic modes is still much greater than $1/2$ but for the latter mode $Q\approx 1.8$. The damping times are much shorter than in the previous environments: $\tau_{-}=\tau_{+,1} \approx 10^5 \ \Rm{s}$ and $\tau_{+,2} \approx 3 \times 10^{-5} \ \Rm{s}$. Then, for $k_{x}c_{\Rm{A}}/\Omega_{p}>0.1$, the damping time of all modes is lower than $\tau=0.01 \ \Rm{s}$, which means that the high frequency ion cyclotron and whistler waves are extremely short-lived in the upper chromosphere. Also, from Equation (\ref{eq:lambda_min}) we obtain a critical wavelength of $\lambda_{c} \approx 40 \ \Rm{m}$, a value that is below the spatial resolution of any currently available instrument. We must note that, although in coronal and solar wind conditions the presence of neutrals is residual, a significant amount of neutrals is present in the chromosphere. Therefore, the results given here for chromospheric conditions should be interpreted with caution because the interactions with neutrals have not been included. The study of this presumably important effect will be tackled in a forthcoming paper. 

	Once we have obtained the solutions of the dispersion relations, we can go back to Equations (\ref{eq:vs+-}) and (\ref{eq:magnetic+-}) and compute the amplitudes of the perturbations associated to each mode. Since we are studying the linear regime, those amplitudes will be proportional to an arbitrary constant. To get rid of that inconvenient factor we focus on the amplitude ratios and their corresponding phase shifts, denoted by $\varphi$. 
	\begin{figure} [ht]
				\centering
				\includegraphics{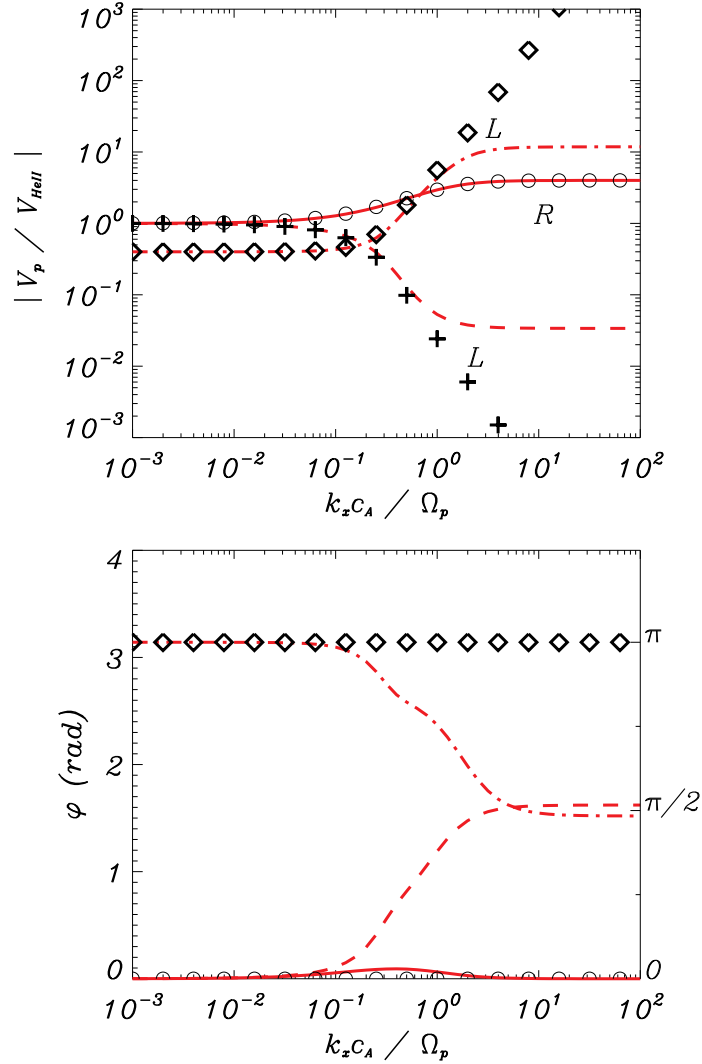}
%		\plotone{2ions_chromosphere_amps.eps}
		\caption{Ratio of amplitudes (top) and phase shift (bottom) of the velocities of ions computed from the solutions of Equations (\ref{eq:dr_2ions_a}) (red lines, with the same style code as in Figure \ref{fig:2ions_wR_wI}) and (\ref{eq:dr_2ions_b}) (black symbols) for a two-ion plasma with upper chromospheric conditions. Note that the black crosses are not shown in the bottom panel: the reason is that they would overlap the circles.}
		\label{fig:amplitudes_impulsive}
	\end{figure}
	
	Figure \ref{fig:amplitudes_impulsive} shows the amplitude ratios (top panel) and the phase shifts (bottom) associated to the solutions of the dispersion relations computed with upper chromospheric conditions. We see that the low-frequency waves (the red solid and dashed lines at low wavenumbers when $\nu_{st} \ne 0$, or the black circles and crosses when $\nu_{st}=0$) have an amplitude ratio $|V_{p}/V_{\Rm{He}\textsc{ii}}| \approx 1$ and a phase shift that is close to zero. This means that for the given frequencies the magnetic field is able to keep the two ionized fluids strongly coupled. There is almost no velocity drift and, hence, the momentum transfer, computed through Equation (\ref{eq:rterms}) and that leads to the damping of the oscillations, can be neglected even when collisions are considered.
	
	At higher frequencies, the different inertia and charge number of each species causes them to have unlike responses to the perturbations and the interaction through the magnetic field is not enough to maintain the strong coupling. Thus, velocity drifts appear and, consequently, there is a friction force. The larger the phase shift, the larger the damping caused by the momentum transfer between ions. The modes with larger phase shifts are left-handed polarized. 
	
	If we replace the parameters of the chromosphere with those for the solar corona or the solar wind, the amplitude ratios and phase shifts of waves are well described by the collisionless results of Figure \ref{fig:amplitudes_impulsive}. In those cases, for high-frequency waves there are larger differences between the velocity amplitudes of the two species, except for the right-handed mode.  

\subsubsection{Periodic driver} \label{sec:periodic_2ions}
	Here we focus on the study of waves excited by a periodic driver. We assume a real frequency and solve Equations (\ref{eq:dr_2ions_a}) and (\ref{eq:dr_2ions_b}) as functions of $k_{x}$. The wavenumber may be complex and can be written as $k_{x}=k_{R}+ik_{I}$ (where we drop the subscript ``$x$'' for simplicity). When $\omega >0$, $k_{R}>0$ corresponds to a wave propagating along the positive $x$-axis, while $k_{R}<0$ corresponds to a wave propagating to the opposite direction. When $k_{I} \ne 0$, we find that $\Rm{sgn}(k_{I})=\Rm{sgn}(k_{R})$, meaning that the amplitudes of the perturbations are damped in space.

	Before solving the dispersion relations we can retrieve some of their properties by simple inspection. For instance, Equation (\ref{eq:dr_2ions_b}), which does not take collisions into account, can be written as
	\begin{equation} \label{eq:dr_2ions_b_kx}
		k_{x,\pm}^2=\frac{\omega_{\pm}^2}{c_{\Rm{A}}^2}\frac{\left(1 \mp \frac{\omega_{\pm}}{\widetilde{\Omega}}\right)}{\left(1 \mp \frac{\omega_{\pm}}{\Omega_{1}}\right)\left(1 \mp \frac{\omega_{\pm}}{\Omega_{2}}\right)},
	\end{equation}
	and it can be seen that $k_{x,+}$ has got singular points at $\omega_{+}=\Omega_{1}$ and $\omega_{+}=\Omega_{2}$, while $k_{x,-}$ presents singularities for $\omega_{-}=-\Omega_{1}$ and $\omega_{-}=-\Omega_{2}$. These singularities are known as ion cyclotron resonances \citep[see, e.g.,][]{2001paw..book.....C,2010Rahbarnia}. In such points, the wavenumber tends to infinity and the phase speed goes to zero, meaning that the perturbation does not propagate.

	Conversely, if collisions are considered, Equation (\ref{eq:dr_2ions_a}) can be expressed as
	\begin{equation} \label{eq:dr_2ions_a_kx}
		k_{x,\pm}^2=\frac{\omega_{\pm}^2}{c_{\Rm{A}}^2}\frac{\left(1\mp \frac{\omega_{\pm}}{\widetilde{\Omega}} \mp i\frac{\alpha_{12}}{\widetilde{\Omega}}\frac{\rho_{1}+\rho_{2}}{\rho_{1} \rho_{2}}\right)}{\left(1 \mp \frac{\omega_{\pm}}{\Omega_{1}}\right)\left(1 \mp \frac{\omega_{\pm}}{\Omega_{2}}\right)+ i \Gamma}.
	\end{equation}
	Now there are no singularities. This statement can be proved by equating the denominator to zero and checking that there is no real $\omega$ that solves the resulting equation. The denominator can be expanded in the form of a second degree polynomial on $\omega$; if we compute its discriminant, we find that it is a complex number, which means that there are no real solutions of the equation.
	\begin{figure} [ht]
				\centering
				\includegraphics{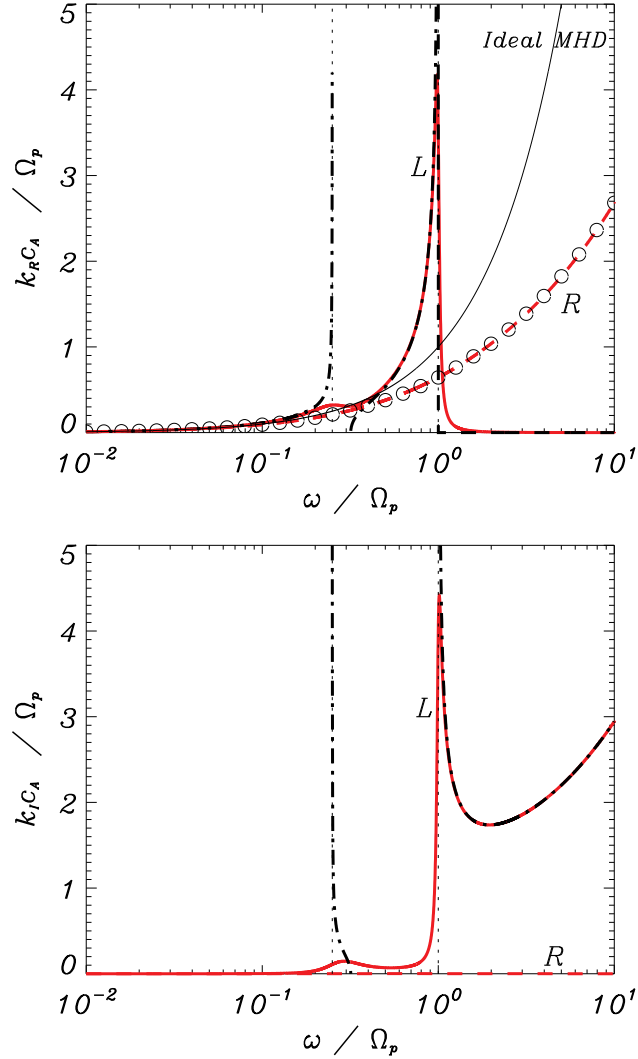}
%		\plotone{2ions_chromosphere_periodic.eps}
		\caption{Normalized wavenumber (top), $k_{R}c_{\Rm{A}}/\Omega_{p}$, and normalized spatial damping (bottom), $k_{I}c_{\Rm{A}}/\Omega_{p}$, as functions of the normalized frequency, $\omega/\Omega_{p}$, for waves excited by a periodic driver in a two-ion plasma with upper chromospheric conditions: $n_{p}=10^{17} \ \Rm{m}^{-3}$, $n_{\Rm{He}\textsc{ii}}=10^{16} \ \Rm{m}^{-3}$, $B_{0}=35 \ \Rm{G}$, $T_{p}=T_{\Rm{He}\textsc{ii}}=10^4 \ \Rm{K}$, and $\nu_{p\Rm{He}\textsc{ii}}=8500 \ \Rm{Hz}$. Red solid lines and red dashed lines correspond to the L and R modes, respectively, when the effect of collisions is included. The black dot-dashed lines and the black circles represent the collisionless left-hand and right-hand modes, respectively. The dotted vertical lines show the position of the resonances and the thin black lines represent the solutions from ideal MHD.}
		\label{fig:2ions_resonances}
	\end{figure}
	
	In Figure \ref{fig:2ions_resonances} we show the study of the effect of collisions on this kind of waves and compare the single-fluid and the multi-fluid models for upper chromospheric conditions. The top panel displays the real part of the normalized wavenumber, $k_{R}c_{\Rm{A}}/\Omega_{p}$, as a function of the normalized frequency, $\omega/\Omega_{p}$, and the bottom panel displays the corresponding imaginary part or normalized spatial damping of the waves, $k_{I}c_{\Rm{A}}/\Omega_{p}$. From the dispersion relations we obtain two solutions for each state of polarization, but we only represent those with $k_{R}>0$, for the sake of simplicity.

	At low frequencies there are no remarkable differences between the cases with $\nu_{p\Rm{He}\textsc{ii}}=0$ and with $\nu_{p\Rm{He}\textsc{ii}} \ne 0$. In this limit, the two circularly polarized modes share the same wavenumber, which coincides with the ideal MHD result. But they start separating from each other when $\omega/\Omega_{p}\gtrsim 0.1$.
		
	If collisions are neglected, the wavenumber of the L mode rises very fast until it reaches a first resonance in $\omega/\Omega_{p}=0.25$ (or equivalently, $\omega=\Omega_{\Rm{He}\textsc{ii}}$). Then, it enters a cut-off region where $k_{R}=0$ and $k_{I}>0$: waves are evanescent. The cut-off region ends when $\omega=\widetilde{\Omega}$ and $k_{R}$ increases again until it finds a second resonance in $\omega=\Omega_{p}$. From this value on, the mode becomes evanescent again. The R mode is not subject to any resonance and its wavenumber keeps increasing with the frequency, but it is always lower than the result provided by ideal MHD.

	In the case where $\nu_{p\Rm{He}\textsc{ii}} \ne 0$, we find that collisions have a very small impact on the R mode. On the contrary, the behavior of the L mode is dramatically altered: the two resonances are removed and $k_{R}$ remains finite. Moreover, the first cut-off region is also removed: there is some damping on the perturbations but they do not turn into evanescent waves. Finally, from $\omega/\Omega_{p}=1$ on, the normalized wavenumber suffers a strong decrease and the waves are then overdamped instead of being fully evanescent.
	\begin{figure} %[ht]
				\centering
				\includegraphics{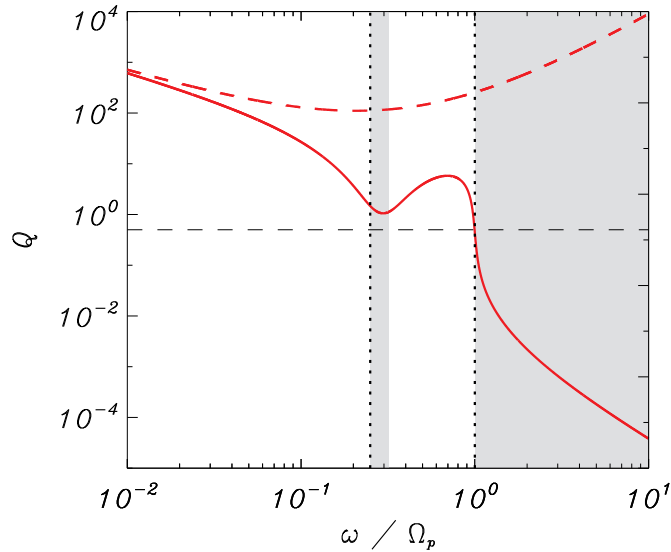}
		%\plotone{2ions_qfactor.eps}
		\caption{Quality factor, $Q$, as a function of the normalized frequency, $\omega/\Omega_{p}$, computed from the solutions displayed on Figure \ref{fig:2ions_resonances}. The shaded areas show the cut-off regions of the left-hand polarized mode from Equation (\ref{eq:dr_2ions_a_kx}). The red solid and red dashed curves represent the left-hand and right-hand modes of Equation (\ref{eq:dr_2ions_b_kx}), respectively. The dotted vertical lines mark the position of the cyclotron frequencies, with $\Omega_{\Rm{He}\textsc{ii}}<\Omega_{p}$. The dashed horizontal line corresponds to $Q=1/2$.}  
		\label{fig:2ions_qfactor}
	\end{figure}
	
	The discussion in the previous paragraph is better illustrated by Figure \ref{fig:2ions_qfactor}, where the quality factor, now defined as $Q\equiv 1/2|k_{R}/k_{I}|$, is shown as a function of the normalized frequency, $\omega/\Omega_{p}$. The shaded areas mark the cut-off regions of the collisionless L mode. The horizontal dashed line points out the critical value $Q=1/2$. It can be seen that the R mode is always underdamped ($Q>1/2$), with a minimum of $Q$ around $\omega/\Omega_{p}=0.25$, i.e., at the frequency of resonance of the singly ionized helium. Waves associated with the L mode are clearly underdamped at low frequencies. When $\omega/\Omega_{p}$ rises, $Q$ decreases until there is a minimum in the first cut-off region; however, even in that region, such waves are still underdamped since $Q\approx 1$. Then, $Q$ increases again and there is a maximum before $\omega/\Omega_{p}=1$, position at which the curve crosses the critical value $Q=1/2$ and oscillations become overdamped. Although $Q \ne 0$ (contrary to what happens when there are no collisions), $Q$ decreases at a very fast rate in the second cut-off region and waves may well be treated as evanescent for very large frequencies.
	
	The removal of resonances and cut-offs due to collisions can be understood in physical terms as follows. It is a consequence of the dissipation caused by the friction between the different species. For instance, in the case without collisions, waves do not propagate at the resonant frequencies and the energy provided by the driver is used in increasing the radius of gyration of the ions. But if there is friction, a fraction of that energy is transferred to the other species and thus the perturbation is allowed to propagate.
	
	The effect of collisions described in the previous paragraphs is consistent with what is shown in Section I of \citet{2010Rahbarnia}, although the plasmas investigated in that paper have a different composition than the ones studied in the present work.

	\begin{figure}
				\centering
				\includegraphics{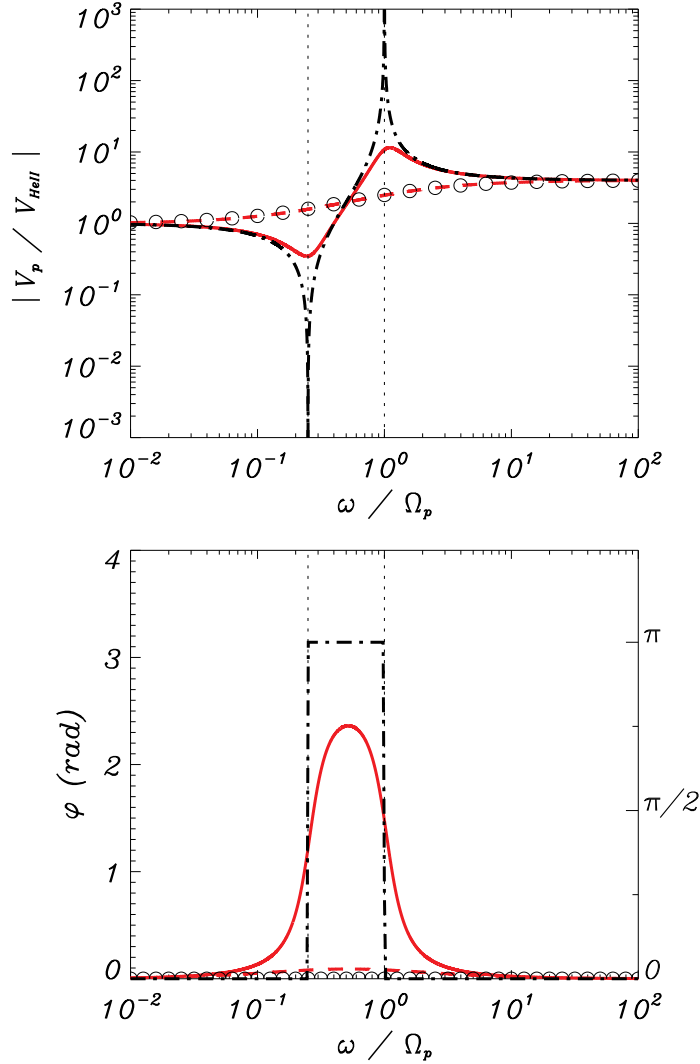}
%		\plotone{2ions_chromosphere_amps_periodic.eps}
		\caption{Ratio of amplitudes (top) and phase shift (bottom) of the velocities of protons and singly ionized helium computed from the results shown on the right panels of Figure \ref{fig:2ions_resonances} (chromospheric conditions).}
		\label{fig:amplitudes_periodic}
	\end{figure}

	Now, we follow the same procedure as in the section dedicated to the impulsive driver and analyze the amplitudes of the perturbations. In the top panel of Figure \ref{fig:amplitudes_periodic} we show the absolute value of the ratio of velocities, $|V_{p}/V_{\Rm{He}\textsc{ii}}|$, for the solutions presented on Figure \ref{fig:2ions_resonances}. In the bottom panel, the phase shift, $\varphi$, is shown. We see that at low frequencies the amplitude ratios are close to unity and that there is no phase shift. This means that the magnetic field produces a strong coupling between the two species. But as the frequency increases, the motions of the two fluids become more independent from each other, specially at the resonances of the L mode. The inclusion of friction causes the fluids to be more coupled. This can be clearly seen at the frequencies of resonance, where the amplitude ratios reach some extreme values when collisions are neglected.

	We do not include figures with the results for the solar corona or the solar wind because, as it has already been mentioned, the effect of collisions is almost negligible in those conditions. Thus, the behavior of waves in those plasmas can be anticipated from the collisionless solutions in Figures \ref{fig:2ions_resonances} and \ref{fig:amplitudes_periodic} but keeping in mind that now the second ionized species is $\Rm{He} \ \textsc{iii}$ instead of $\Rm{He} \ \textsc{ii}$; this change implies that the positions of the lower resonance and the lower bound of the cut-off region are modified accordingly.

\subsection{Waves in a three-ion plasma} \label{sec:dispersion_three-ion}
	We move to the case of a fully ionized plasma composed of three different species. As already mentioned, the complete dispersion relation can be obtained from Equation (\ref{eq:dispersion}), but it is too long and complex to be shown here. Nonetheless, the abridged version in which the elastic collisions between the different ions are ignored allows us to analyze some of the general properties. This simpler collisionless version is given by   
	\begin{eqnarray} \label{eq:dr_3ions_a}
		&&\omega_{\pm}^2\big[Z_{1}n_{1}\left(\omega_{\pm} \mp \Omega_{2}\right)\left(\omega_{\pm} \mp \Omega_{3}\right) \nonumber \\ 
		&+&Z_{2}n_{2}\left(\omega_{\pm} \mp \Omega_{1}\right) \left(\omega_{\pm} \mp \Omega_{3}\right) \nonumber \\ 
		&+&Z_{3}n_{3}\left(\omega_{\pm} \mp \Omega_{1}\right)\left(\omega_{\pm} \mp \Omega_{2}\right)\big] \nonumber \\
		&\pm& \frac{B_{x}k_{x}^2}{e \mu_{0}}\left(\omega_{\pm} \mp \Omega_{1}\right)\left(\omega_{\pm}\mp \Omega_{2}\right)\left(\omega_{\pm} \mp \Omega_{3}\right)=0.
	\end{eqnarray}

	These new dispersion relations are fourth degree polynomials in $\omega$, so for the case of waves generated by an impulsive driver we find one additional oscillation mode for each polarization compared to the system with only two ions. Repeating the procedure of Section \ref{sec:impulsive_2ions}, the limit of low wavenumber and frequencies may be investigated analytically. It can be checked that the new mode is related to the cyclotron frequencies, instead of the Alfvén frequency: each polarization still has only two Alfvénic modes, with $\omega \approx \pm \omega_{\Rm{A}}$, where the Alfvén speed is computed using the sum of the densities of the three ions; the frequencies of the remaining modes are $\omega_{\pm}= \pm \widetilde{\Omega_{1}}$ and  $\omega_{\pm}= \pm \widetilde{\Omega_{2}}$, where $\widetilde{\Omega_{1}}$ and $\widetilde{\Omega_{2}}$ are the solutions to
	\begin{eqnarray} \label{eq:3ions_kx0}
		&& Z_{1}n_{1}\left(\omega-\Omega_{2}\right)\left(\omega-\Omega_{3}\right) + Z_{2}n_{2} \left(\omega-\Omega_{1}\right)\left(\omega-\Omega_{3}\right) \nonumber \\
		&+& Z_{3}n_{3}\left(\omega-\Omega_{1}\right)\left(\omega-\Omega_{2}\right)=0,
	\end{eqnarray}
	and are given in the Appendix \ref{app:B} by Equations (\ref{eq:weighted_1}) and (\ref{eq:weighted_2}), respectively. We note that no additional solution appears in the case of a periodic driver, although a third resonance is present when $\omega_{\pm}=\pm \Omega_{3}$. If the limit $n_{3} \to 0$ is taken, it is easy to verify that the formulas for the system with only two ions are recovered.

	Abundances of $p$ and $\Rm{He} \ \textsc{iii}$ in the solar corona and the solar wind are much larger than the abundances of any other ions \citep[see, e.g.,][]{1977SoPh...53..409A,1989GeCoA..53..197A}. Hence, the addition of a third ion would hardly modify the results from the two-ion model when applied to those two environments. However, looking at the model F of \citet{1993ApJ...406..319F} we find that for a height of $\sim 2016 \ \Rm{km}$ over the top of the photosphere the number densities are $n_{p}\approx 7 \times 10^{16} \ \Rm{m^{-3}}$, $n_{\Rm{He}\textsc{ii}}\approx 6 \times 10^{15} \ \Rm{m^{-3}}$, and $n_{\Rm{He}\textsc{iii}} \approx 10^{15} \ \Rm{m^{-3}}$; at such height the contribution of the three ions should be considered, although protons are still the dominant species. With these values, the corresponding temperature of $T\approx 2 \times 10^{4} \ \Rm{K}$ and a magnetic field $B_{x}=35 \ \Rm{G}$, we obtain the following collision and cyclotron frequencies: $\nu_{p\Rm{He}\textsc{ii}} \approx 2000 \ \Rm{Hz}$, $\nu_{\Rm{He}\textsc{ii}p} \approx 5840 \ \Rm{Hz}$,  $\nu_{p\Rm{He}\textsc{iii}} \approx 1260 \ \Rm{Hz}$, $\nu_{\Rm{He}\textsc{iii}p} \approx 22100 \ \Rm{Hz}$, $\nu_{\Rm{He}\textsc{ii}\Rm{He}\textsc{iii}} \approx 540 \ \Rm{Hz}$, $\nu_{\Rm{He}\textsc{iii}\Rm{He}\textsc{ii}} \approx 3250 \ \Rm{Hz}$, $\Omega_{p}=335268 \ \Rm{rad \ s^{-1}}$, $\Omega_{\Rm{He}\textsc{ii}}=83817.1 \ \Rm{rad \ s^{-1}}$, and $\Omega_{\Rm{He}\textsc{iii}}=167634 \ \Rm{rad \ s^{-1}}$. The Alfvén speed is $c_{\Rm{A}} \approx 244 \ \Rm{km \ s^{-1}}$. 
	
\begin{figure}
					\centering
					\includegraphics{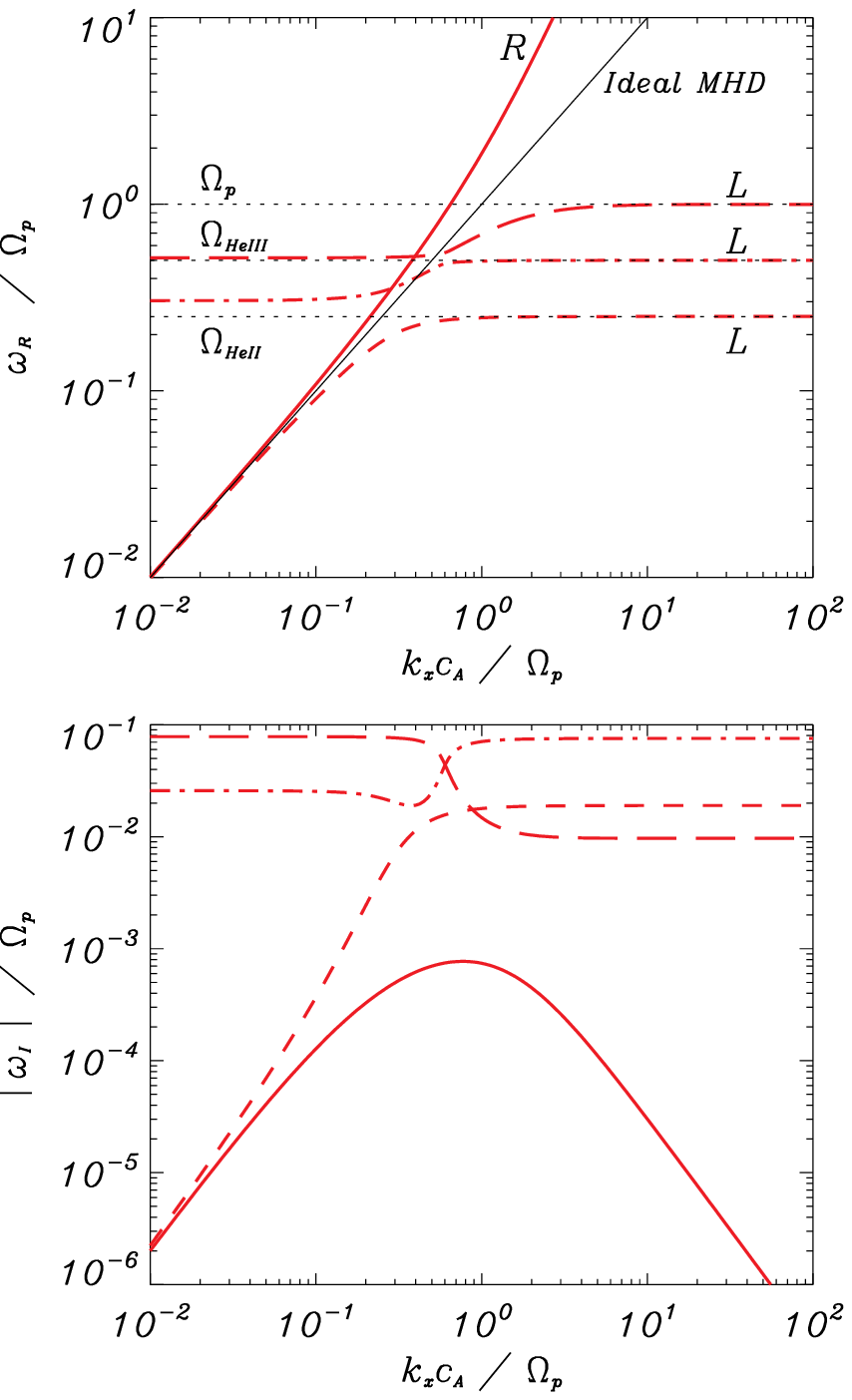}
%		\plotone{3ions_chromosphere.eps}
		\caption{Solutions to the dispersion relations of a three-ion plasma with $n_{p}=7 \times 10^{16} \ \Rm{m^{-3}}$, $n_{\Rm{He}\textsc{ii}}=6 \times 10 ^{15} \ \Rm{m^{-3}}$, $n_{\Rm{He}\textsc{iii}}=10^{15} \ \Rm{m}^{-3}$, $B_{x}=35 \ \Rm{G}$, $\nu_{p\Rm{He}\textsc{ii}} \approx 2000 \ \Rm{Hz}$, $\nu_{\Rm{He}\textsc{ii}p} \approx 5840 \ \Rm{Hz}$,  $\nu_{p\Rm{He}\textsc{iii}} \approx 1260 \ \Rm{Hz}$, $\nu_{\Rm{He}\textsc{iii}p} \approx 22100 \ \Rm{Hz}$, $\nu_{\Rm{He}\textsc{ii}\Rm{He}\textsc{iii}} \approx 540 \ \Rm{Hz}$, and $\nu_{\Rm{He}\textsc{iii}\Rm{He}\textsc{ii}} \approx 3250 \ \Rm{Hz}$. Top: normalized real part of the frequency as a function of the normalized wavenumber. Bottom: absolute value of the normalized damping as a function of $k_{x}c_{\Rm{A}}/\Omega_{p}$. Red dashed lines represent the L modes and the red solid line represents the R mode. The black thin line corresponds to the solution of ideal MHD.}
		\label{fig:3ions_impulsive}
	\end{figure}

	To analyze the properties of waves generated by an impulsive driver we insert the parameters into Equations (\ref{eq:dispersion}) and (\ref{eq:dr_3ions_a}) and study their dependence on a real wavenumber, $k_{x}$. The results are shown in Figure \ref{fig:3ions_impulsive}, where once more the solutions to the collisionless dispersion relation are not plotted because there are no appreciable differences with Equation (\ref{eq:dispersion}) in the real part of the frequency (top panel) and $\omega_{I}=0$. Again only the solutions with $\omega_{R}>0$ are displayed.

	In the limit of low wavenumbers, we verify that two of the solutions coincide with the Alfvén frequency provided by the single-fluid description while the other two are given by the values $\widetilde{\Omega_{1}}$ and $\widetilde{\Omega_{2}}$ reported in the previous paragraphs. When $k_{x}c_{\Rm{A}}/\Omega_{p}$ increases, the Alfvénic L mode (dashed line) turns into an ion cyclotron mode and its frequency tends to the lower cyclotron frequency (the same behavior that is found in the two-ion description). The remaining L modes tend to the limiting values $\Omega_{\Rm{He}\textsc{iii}}$ and $\Omega_{p}$ and they conserve their order: the mode associated with $\widetilde{\Omega_{2}}$ (which is larger than $\widetilde{\Omega_{1}}$) tends to the upper cyclotron frequency. In the last place, the Alfvénic R mode becomes the whistler wave and its frequency is always higher than the Alfvén frequency.
	
	The bottom panel shows us that the R mode is again the less affected by collisions and its normalized damping is $|\omega_{I}|/\Omega_{p} < 10^{-5}$ for very low and very high wavenumbers, with a maximum of $|\omega_{I}|/\Omega_{p}\approx 10^{-3}$ around $\omega_{\Rm{A}}=\Omega_{p}$. For low wavenumbers, the solutions with a bigger damping are those related to $\widetilde{\Omega_{\Rm{1}}}$ and $\widetilde{\Omega_{2}}$: the damping of the former (dot-dashed line) decreases with the wavenumber until it reaches a minimum around $\omega_{\Rm{A}}=0.5 \Omega_{p}$, then increases again and becomes constant for very large wavenumbers, with $|\omega_{I}|/\Omega_{p}\approx 0.075$; the damping of the mode associated with $\widetilde{\Omega_{2}}$ decreases very fast in the region around $\omega_{\Rm{A}}=\Omega_{p}$ and then stabilizes in $|\omega_{I}|/\Omega_{p}\approx 0.025$. Finally, the damping of the Alfvénic L mode increases with the wavenumber until it reaches a value $|\omega_{I}|/\Omega_{p} \approx 0.06$ when $\omega_{\Rm{A}}/\Omega_{p}>1$.
	
	\begin{figure}
		\centering
		\includegraphics{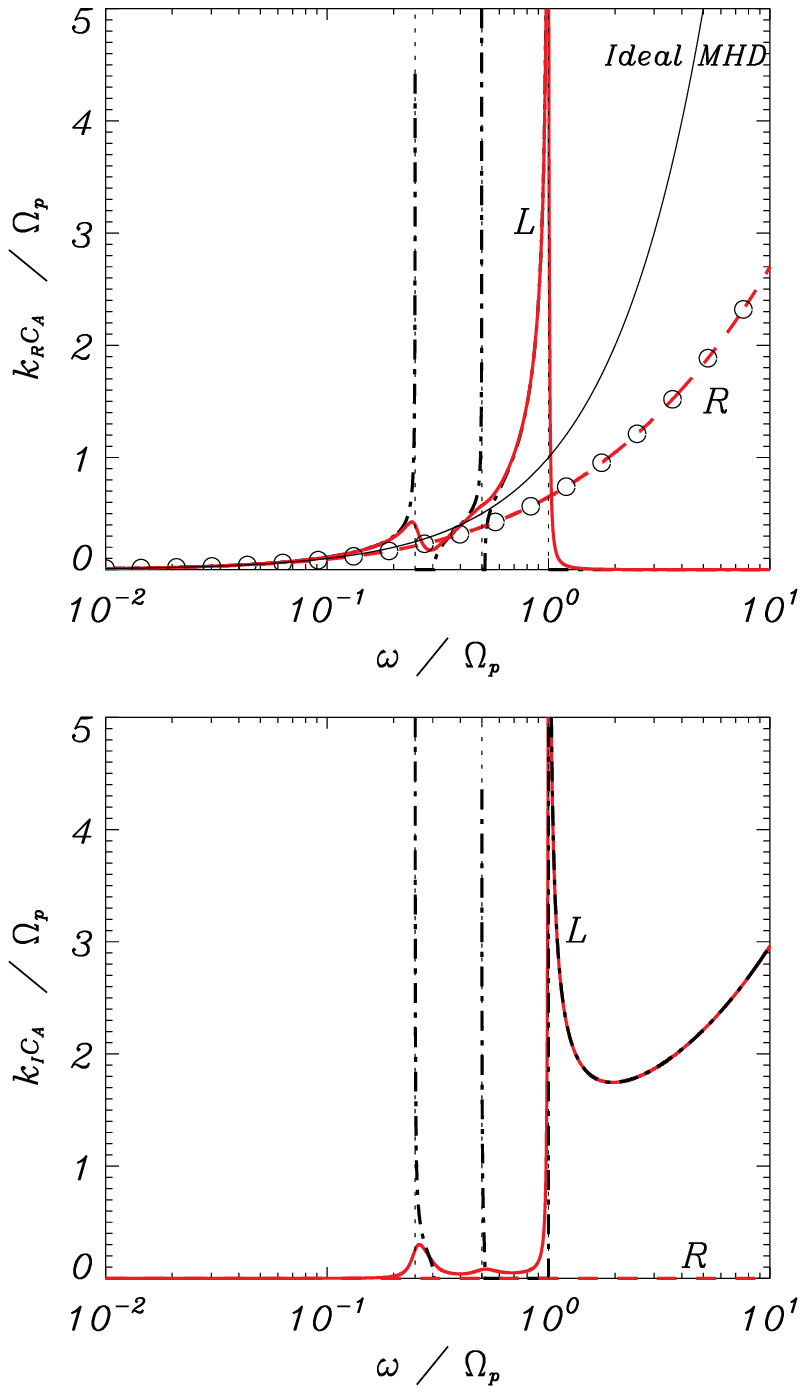}
		%		\plotone{3ions_chromosphere_periodic.eps}
		\caption{Solutions of the dispersion relations for waves generated by a periodic driver in a three-ion plasma with the same parameters as those used in Figure \ref{fig:3ions_impulsive}.}
		\label{fig:3ions_periodic}
	\end{figure}

	Then we turn to the study of properties of waves excited by a periodic driver, so we investigate the dependence of the dispersion relations on a real frequency, $\omega$. Figure \ref{fig:3ions_periodic} displays the results of this investigation but only the solutions with $k_{\Rm{R}}>0$ are plotted. It can be noticed that the collisionless L mode (black dot-dashed line) exhibits the expected three resonances ($k_{x} \to \infty$) at $\omega=\Omega_{\Rm{He}\textsc{ii}}$, $\omega=\Omega_{\Rm{He}\textsc{iii}}$ and $\omega=\Omega_{p}$. It also has three cut-off regions (instead of the two that exist according to the two-ion model) in which the waves become evanescent: $k_{R}=0$ and $k_{I}\ne 0$ if $\omega \in \left(\Omega_{\Rm{He}\textsc{ii}},\widetilde{\Omega_{\Rm{1}}}\right)$, $\omega \in \left(\Omega_{\Rm{He}\textsc{iii}},\widetilde{\Omega_{\Rm{2}}}\right)$ or $\omega > \Omega_{p}$, where $\widetilde{\Omega_{1}} \approx 0.3 \Omega_{p}$ and $\widetilde{\Omega_{2}} \approx 0.52 \Omega_{p}$. If $\nu_{st} \ne 0$, the singularities are substituted by extrema in the normalized wavenumber, where the highest peak corresponds to the most abundant species, i.e., protons. Again the momentum transfer removes the cut-off regions. For the R mode we find the same behavior that was explained in Section \ref{sec:periodic_2ions}: it has no resonances, its normalized wavenumber increases with the frequency and the spatial damping is $k_{I}c_{\Rm{A}}/\Omega_{p}\ll 1$ at any $\omega/\Omega_{p}$.

	Therefore, the overall results obtained in the three-ion model appear as natural extensions to the results of the two-ion case. Hence, the generalization to plasmas with a larger number of ions is straightforward.

\section{Numerical simulations} \label{sec:simulations}
	In the previous section, through the examination of analytic dispersion relations, we have learnt the properties of the linear wave modes that are present in a multi-ion plasma. Here, we go a step forward and explore the wave behavior from the point of view of numerical simulations. This provides us with some insight on the time-dependent evolution of the waves. In addition, the comparison between analytic and numerical results allows us to test the numerical code.
	
	We use the numerical code MolMHD, based on the method of lines \citep{1963SarminChudov,schiesser1991numerical}, to compute the temporal evolution of Equations (\ref{eq:cont_s})-(\ref{eq:pres_s}), together with Faraday's law, Equation (\ref{eq:induction}). Spatial derivatives are computed with a $4^{th}$ order of accuracy central finite differences scheme while the temporal variable is advanced through a explicit $3^{th}$ degree TVD Runge-Kutta method, where TVD refers to total variation diminishing \citep{Harten:1983:HRS}. We note that the CFL condition \citep{1928MatAn.100...32C} imposes a strong constraint to the maximum time step that can be used in the simulations. This is mainly due to the presence of the ion cyclotron frequencies, but also to the diffusion scales related to collisions.

	Since we aim to compare the outcome of the numerical simulations with the results provided by the dispersion relations examined in the previous section, we perform 1D simulations with an initially uniform and static medium, so $\rho_{s}(x)=\rho_{s,0}$ and $\bm{V_{s,0}}\left(x\right)=0$. The background magnetic field is given by $\bm{B_{0}}\left(x\right)=\left(B_{x},0,0\right)$. Here, we are interested in the linear regime of incompressible perturbations that are transverse to the $x$-direction. However, we note that the numerical code is nonlinear so that it consistently computes nonlinear effects. The study of nonlinearities is left for a future paper of the series beyond this initial and exploratory study. In Section \ref{sec:sims_impulsive} we analyze waves excited by an impulsive driver and in Section \ref{sec:sims_periodic} we present the study of the case with a periodic driver.

\subsection{Impulsive driver} \label{sec:sims_impulsive}
	Here we describe the simulations we perform to investigate standing waves excited by an impulsive driver. We use a uniform grid of $N=401$ points to cover the domain $x \in [-L,L]$, where $L$ is a length scale, and apply boundary conditions that impose that the velocity perturbations are equal to zero at $x=\pm L$. The initial conditions we consider are
	\begin{equation} \label{eq:init_cond}
		\bm{V_{s}}(t=0)= \left( \begin{array}{c}
		0 \\
		A_{s,y}\cos \left(k_{x}x\right) \\ 
		A_{s,z}\cos \left(k_{x}x\right)
		\end{array} \right),
	\end{equation}
	where the wavenumber is $k_{x}=\pi/2L$. This initial condition corresponds to the fundamental standing wave in the closed domain. There is no initial perturbation of the remaining variables (magnetic field, densities and pressures). To simplify the analysis we set $A_{s,z}=0$, so that the initial perturbation has only a $y$-component in velocity. Due to the symmetry of the system, the results would be equivalent if we chose $A_{s,y}=0$ and $A_{s,z}\ne0$. As we want to focus on the linear regime we use amplitudes of the velocity perturbations that fulfill $A_{s,y}\ll c_{\Rm{A}}$.
	
\subsubsection{Two-ion plasmas} \label{sec:sims_impulsive_2ions}
	To start with, we analyze the case of plasmas made of two different ionized species. We focus on the upper chromospheric region, where the effect of collisions between the two fluids is much more relevant than in the corona or in the solar wind, as it has been shown in Section \ref{sec:dispersion}.
	
\begin{figure}
		\centering
		\includegraphics{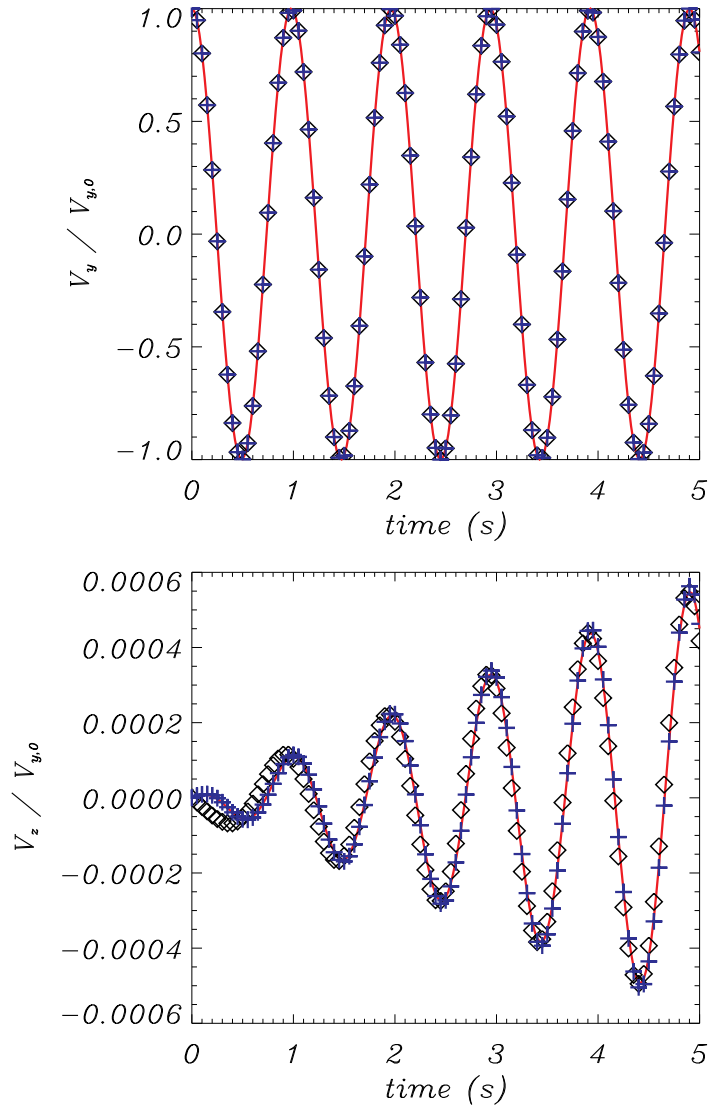}
		%		\plotone{sim_chromosphere_a.eps}
		\caption{Simulation of an Alfvén wave in a two-ion plasma with chromospheric conditions: $n_{p}=10^{17} \ \Rm{m}^{-3}$, $n_{\Rm{He}\textsc{ii}}=0.1n_{p}$, $B_{x}=35 \ \Rm{G}$, and $\nu_{p\Rm{He}\textsc{ii}}=8500 \ \Rm{Hz}$. The initial perturbation is given by Equation (\ref{eq:init_cond_1}) and the wavenumber is $k_{x}=\pi/10^{5} \ \Rm{m}^{-1}$. The top panel shows the normalized $y$-component of the velocity of ions, $V_{y}/V_{y,0}$, at the position $x=0$; the $z$-component, $V_{z}/V_{y,0}$, is shown in the bottom panel. The red lines represent the velocity of protons, the black diamonds represent the velocity of $\Rm{He}\textsc{ii}$ and the blue crosses correspond to the analytic fits given by Equation (\ref{eq:fit}).} 
		\label{fig:sim_2ions_a}
	\end{figure}

	Figure \ref{fig:sim_2ions_a} displays the results of a simulation in which the initial perturbations are
	\begin{equation} \label{eq:init_cond_1}
		V_{p,y}(t=0)=V_{\Rm{He}\textsc{ii},y}(t=0)=10^{-3}c_{\Rm{A}}\cos \left(k_{x}x\right),
	\end{equation}
	and $k_{x}=\pi/10^{5} \ \Rm{m}^{-1}$. The top panel displays the $y$-component of the velocity of protons (solid red line) and singly ionized helium (black diamonds) at the position $x=0$. The bottom panel shows the respective $z$-components. As the simulation stays in the linear regime, the values of the amplitudes of the perturbations are not important, only the ratios between those magnitudes are relevant. Therefore, we have normalized the results with respect to the initial amplitude of the $y$-component of the velocity of protons, $V_{y,0}\equiv V_{p,y}(t=0)$.

	On the top panel it can be seen that the $y$-components of the velocities of the two ions are strongly coupled: they oscillate with the same frequency, amplitude and phase. On the contrary, the z-components show some differences in the first steps of the simulation. The $\Rm{He} \ \textsc{ii}$ fluid starts to oscillate with a phase shift with respect to the protons, but as time increases this phase shift is reduced. In addition, the amplitude of the $z$-component initially is much smaller than that of the $y$-component. This is due to the fact that the wave we are investigating is a combination of various modes of oscillation. From any of the two panels it can be checked that both ions oscillate with a frequency $\omega \approx 6.41 \ \Rm{rad \ s^{-1}}$, which coincides with the low frequency solutions obtained from Equation (\ref{eq:dr_2ions_a}). The solutions from the multi-fluid model that are associated with the cyclotron frequencies are not found in this simulation. This absence may be due to an insufficient temporal resolution or it may be caused by the specific choice of the initial conditions. This issue is investigated later.

	The simulated wave may be a combination of various modes predicted by the dispersion relation. The velocity perturbation associated with each mode can be written as
	\begin{equation} \label{eq:vs_polarized}
		\bm{V_{s,\pm}}= \left( \begin{array}{c}
		0 \\
		V_{y} \\ 
		\pm iV_{z}
		\end{array} \right)= \left( \begin{array}{c}
		0 \\
		V_{\Rm{0}}\cos \left(\omega_{\pm}t\right) \\ 
		\mp V_{\Rm{0}}\sin \left(\omega_{\pm}t\right)
		\end{array} \right),
	\end{equation}
	which is equivalent to Equation (\ref{eq:polarized}). From Equation (\ref{eq:dr_2ions_a}) we get six modes: three of them are left-hand polarized and the other three are right-hand polarized. But, as already mentioned, in the simulation of Figure \ref{fig:sim_2ions_a} two of them cannot be found. Hence, the oscillation may be expressed as
	\begin{eqnarray} \label{eq:vs_polarized2}
		V_{s,y}&=&A_{s,0}\big[\cos \left(\omega_{+,1}t\right) + \cos \left(\omega_{+,2}t\right) \nonumber \\
		&+& \cos \left(\omega_{-,1}t\right) + \cos \left(\omega_{-,2}t\right)\big], \nonumber \\
		V_{s,z}&=&A_{\Rm{s,0}}\big[-\sin \left(\omega_{\Rm{+,1}}t\right) - \sin \left(\omega_{+,2}t\right) \nonumber \\
		&+& \sin \left(\omega_{-,1}t\right) + \sin \left(\omega_{-,2}t\right)\big],
	\end{eqnarray}
	where $\omega_{\pm,1}$ and $\omega_{\pm,2}$ are the roots of the dispersion relation (we note that these roots have a negligible imaginary part for the chosen parameters). In the most general case, each oscillation mode has a different amplitude, but here we assume that the four amplitudes are equal, $A_{s,0}$, which is consistent with the analysis in Section \ref{sec:impulsive_2ions}. Furthermore, since we have found that $\omega_{-,1}=-\omega_{+,1}$ and $\omega_{-,2}= -\omega_{+,2}$, Equation (\ref{eq:vs_polarized2}) can be rewritten as
	\begin{equation} \label{eq:vs_polarized3}
		\bm{V_{s}}=\left( \begin{array}{c}
		2A_{s,0}\bigl[\cos \left(\omega_{+,1}t\right) + \cos \left(\omega_{+,2}t\right)\bigr] \\
		-2A_{s,0}\left[\sin \left(\omega_{+,1}t\right) + \sin \left(\omega_{+,2}t\right)\right]	
		\end{array} \right),
	\end{equation}
	or, equivalently,
	\begin{equation} \label{eq:fit}
		\bm{V_{s}}=\left( \begin{array}{c}
		4A_{s,0}\cos\big(\frac{\omega_{+,1}+\omega_{+,2}}{2}t\big)\cos\big(\frac{\omega_{+,1}-\omega_{+,2}}{2}t\big) \\
		-4A_{s,0}\sin\big(\frac{\omega_{+,1}+\omega_{+,2}}{2}t\big)\cos\big(\frac{\omega_{+,1}-\omega_{+,2}}{2}t\big)
		\end{array} \right),
	\end{equation}
	which represents the composition of a carrier wave with frequency $\theta_{C}=\left(\omega_{+,1}-\omega_{+,2}\right)/2$ and an envelope wave with frequency $\theta_{E}=\left(\omega_{+,1}+\omega_{+,2}\right)/2$ (we note that $\omega_{+,1}$ and $\omega_{+,2}$ have opposite signs). If we set $A_{s,0}=1/4$, the previous formulas fit very well the velocity of ions, as shown by the blue crosses in Figure \ref{fig:sim_2ions_a}, with the exception of the very first instants.

\begin{figure}
		\centering
		\includegraphics{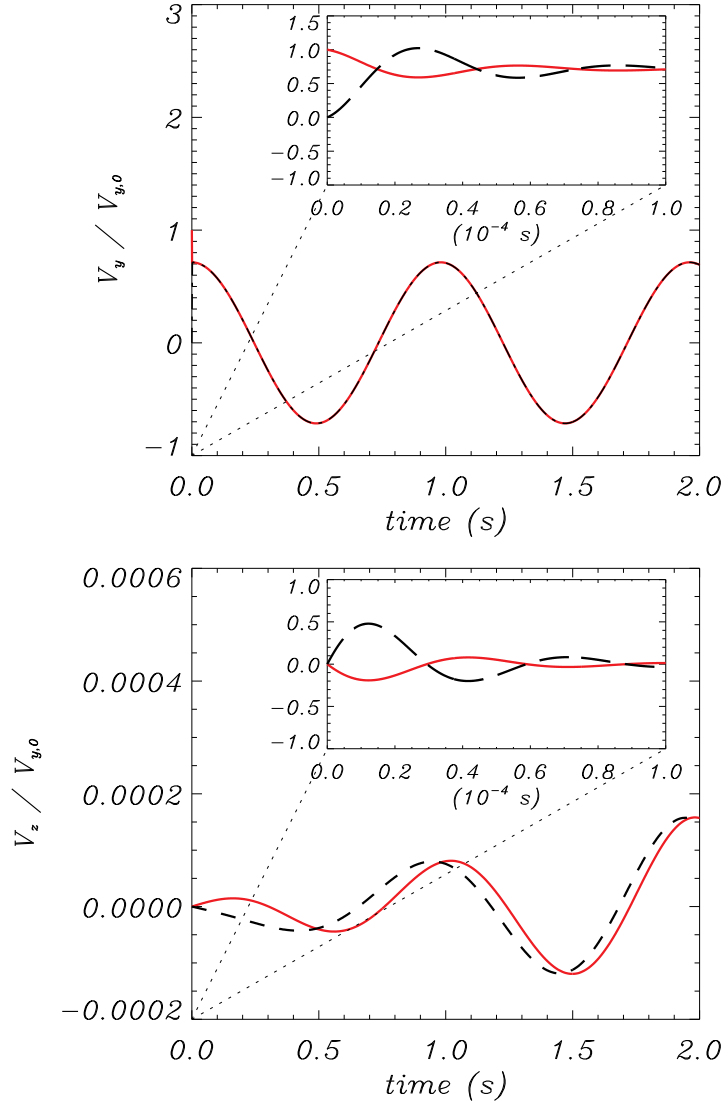} 
		%		\plotone{sim_chromosphere_b.eps}
		\caption{Simulation with the same physical parameters as Figure \ref{fig:sim_2ions_a} (chromospheric conditions) but with a smaller timestep and a different initial perturbation so that the $\Rm{He} \ \textsc{ii}$ fluid is initially at rest. The red solid lines represent the velocity of protons and the black dashed lines represent the velocity of singly ionized helium. The top panel corresponds to the normalized $y$-component of the velocity and the bottom one to the normalized $z$-component.}
		\label{fig:sim_2ions_b}
	\end{figure}

	Now we run a simulation with the same parameters as before but with a different initial perturbation. We set $V_{\Rm{He}\textsc{ii}}(t=0)=0$, i.e., the singly ionized helium fluid is initially at rest. The result of this simulation is plotted in Figure \ref{fig:sim_2ions_b}.  
	It can be seen that after a extremely short relaxation time, the velocities of the two ions become equal and the two species start to oscillate in phase as if they were a single fluid. After performing simulations with different physical parameters and initial conditions, it can be empirically deduced that the amplitude of the oscillation after the relaxation time is given by

	\begin{equation} \label{eq:wave_amp}
		\overline{V}=\frac{\sum_{s}\rho_{s}V_{s}\left(t=0\right)}{\sum_{s}\rho_{s}}.
	\end{equation}

	The insets in Figure \ref{fig:sim_2ions_b} display zooms of the initial time steps of the simulation and reveal the presence of a high frequency oscillation that is damped. We compute the frequency of that oscillation and obtain the value $\omega \approx 106672.7-i29756 \ \Rm{rad \ s^{-1}}$, which agrees with the solution from Equation (\ref{eq:dr_2ions_a}) that could not be found in the previous simulations. Hence, this additional mode is only present in the initial stages of the simulations when the initial velocity amplitudes of the two ions are different. Moreover, we note that the two species oscillate in anti-phase, again consistent with the results from the study of the dispersion relation. If the effect of collisions is ignored, there is no damping and none of the modes should disappear with time.

	\begin{figure*}
		\plottwo{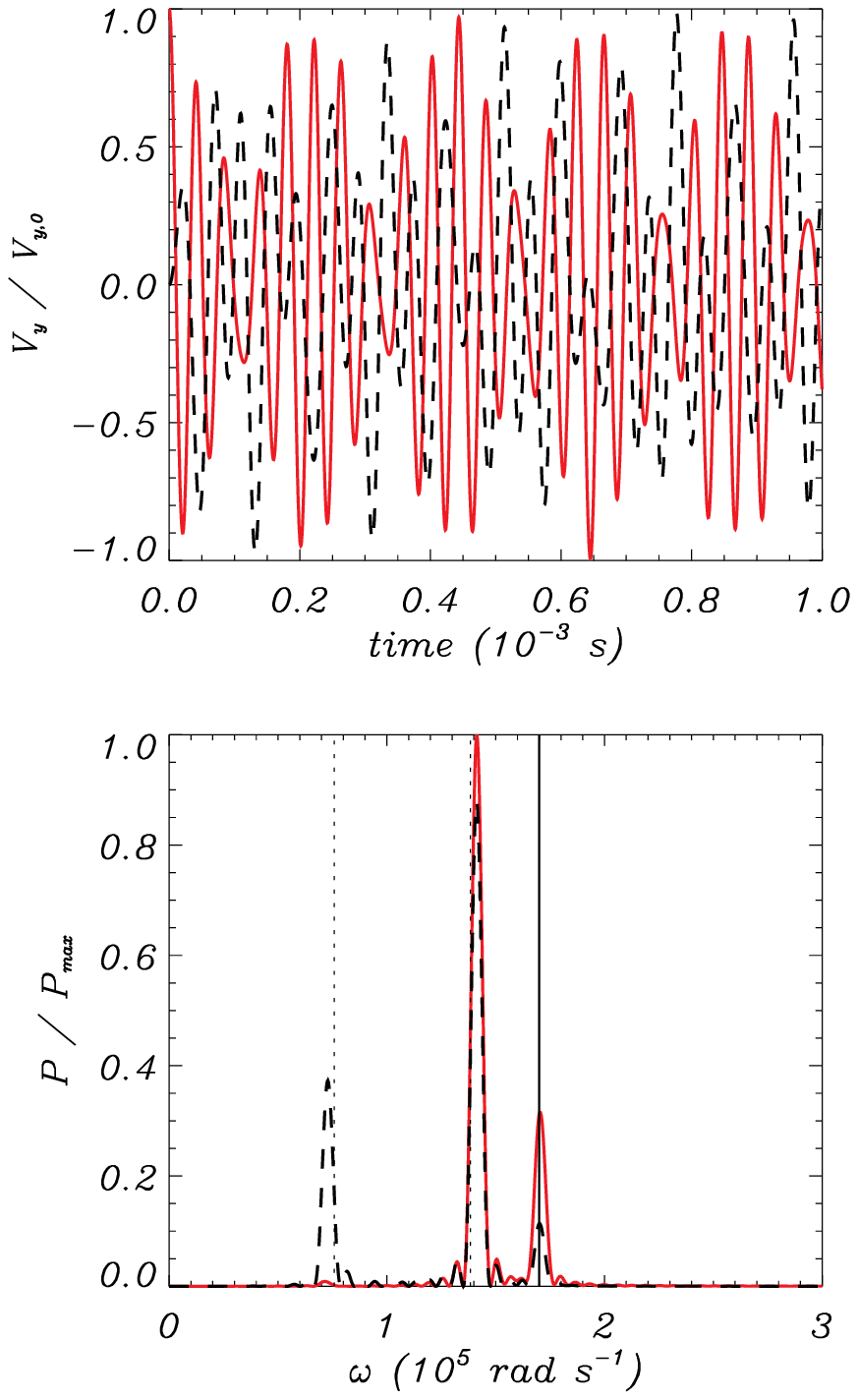}{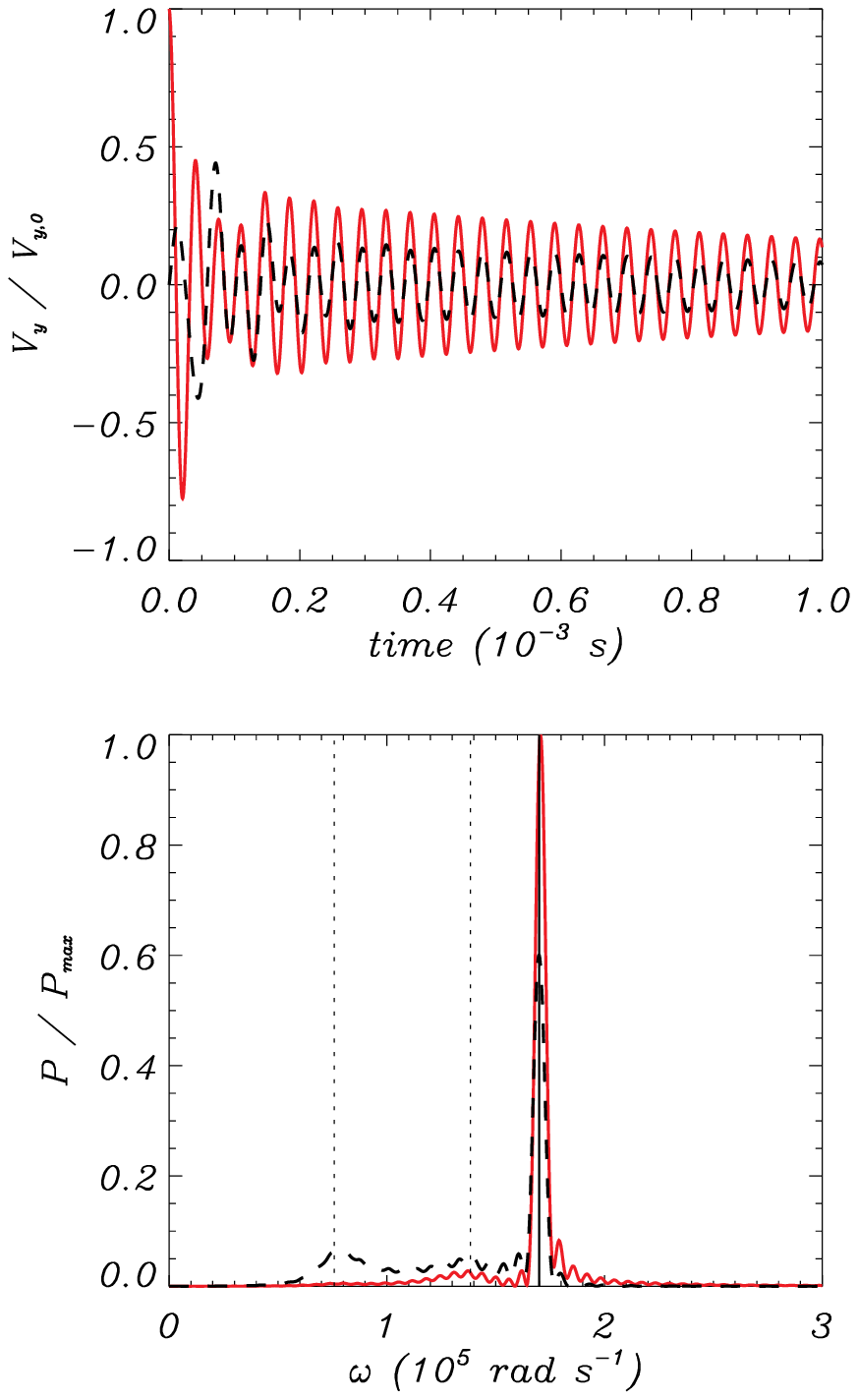}
		\caption{Normalized $y$-component of the velocities of ions (top panels) and spectra of the oscillations (bottom panels) from two different simulations of a two-ion plasma with upper chromospheric conditions. The left and right panels represent the cases without and with collisions, respectively. The wavenumber is $k_{x}=\pi/5 \ \Rm{m^{-1}}$. The vertical lines on the bottom panels show the solutions of the dispersion relation: the dotted lines correspond to the L modes and the solid lines correspond to the R modes.}
		\label{fig:chromosphere_spec}
	\end{figure*}

	The previous simulations correspond to very low values of the wavenumber (i.e., large wavelengths). Now, we focus on higher wavenumbers and set, for instance, $k_{x}=\pi/5 \ \Rm{m^{-1}}$ (so $\omega_{\Rm{A}}/\Omega_{p}\approx 0.38$). We show in Figure \ref{fig:chromosphere_spec} the results of a simulation with the same initial conditions used for Figure \ref{fig:sim_2ions_b}. The left panels correspond to the case where the effect of collisions is ignored, while on the right panels such an effect is taken into account. On the top panel we plot the $y$-component of the velocity of each species. The motions here are more complex than in the case of low wavenumbers. The result of this simulation cannot be easily related with the modes predicted by the dispersion relation. But we can compute the power spectrum and check if it is associated somehow to the solutions of the dispersion relations.

	The power spectra (normalized to the maximum power of the oscillations of protons) are shown on the bottom panel of Figure \ref{fig:chromosphere_spec}, where the vertical lines mark the position of the roots of the dispersion relation. The positions of the peaks in each spectrum inform about the frequencies of the modes that compose the resulting oscillation and their heights show their relative contribution. There are three main peaks that are in very good agreement with the solutions given by the dispersion relation. But it is clear that each mode affects in a different way to each fluid. For instance, when collisions are overlooked, the lowest frequency mode has an important contribution to the oscillation of $\Rm{He} \ \textsc{ii}$ but almost negligible to that of protons. On the contrary, when collisions are considered, the contribution of the L modes is almost negligible compared to the contribution of the whistler mode: the amplitudes of the three modes attenuate very fast with time but the R mode survives for a longer time.
	
\subsubsection{Three-ion plasmas} \label{sec:sims_impulsive_3ions}
	In this section we investigate waves excited by an impulsive driver in the more general case of three-ion plasmas. We use the set of parameters already employed in Section \ref{sec:dispersion_three-ion}, which correspond to a plasma in the upper chromosphere.
	
\begin{figure*}
		\plottwo{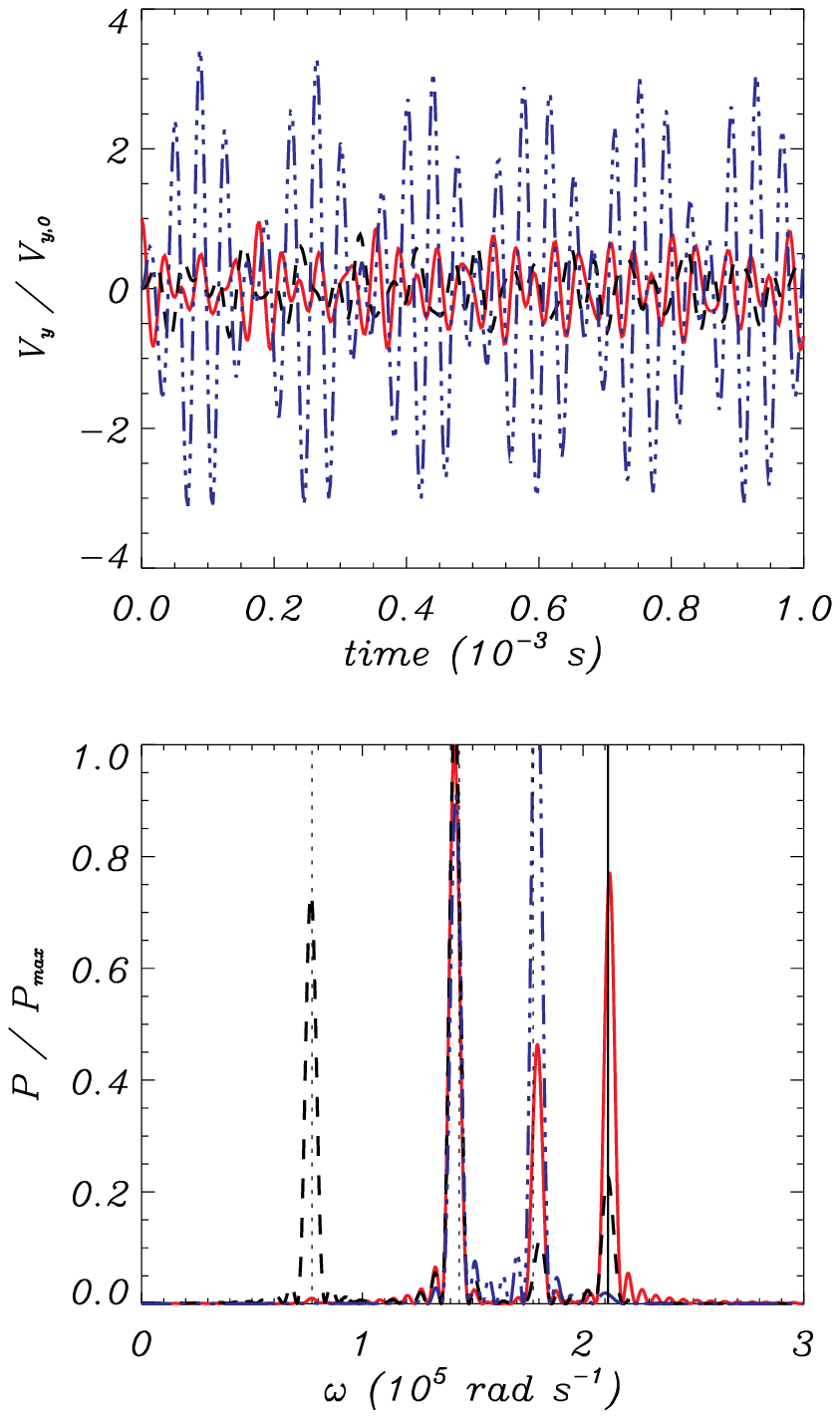}{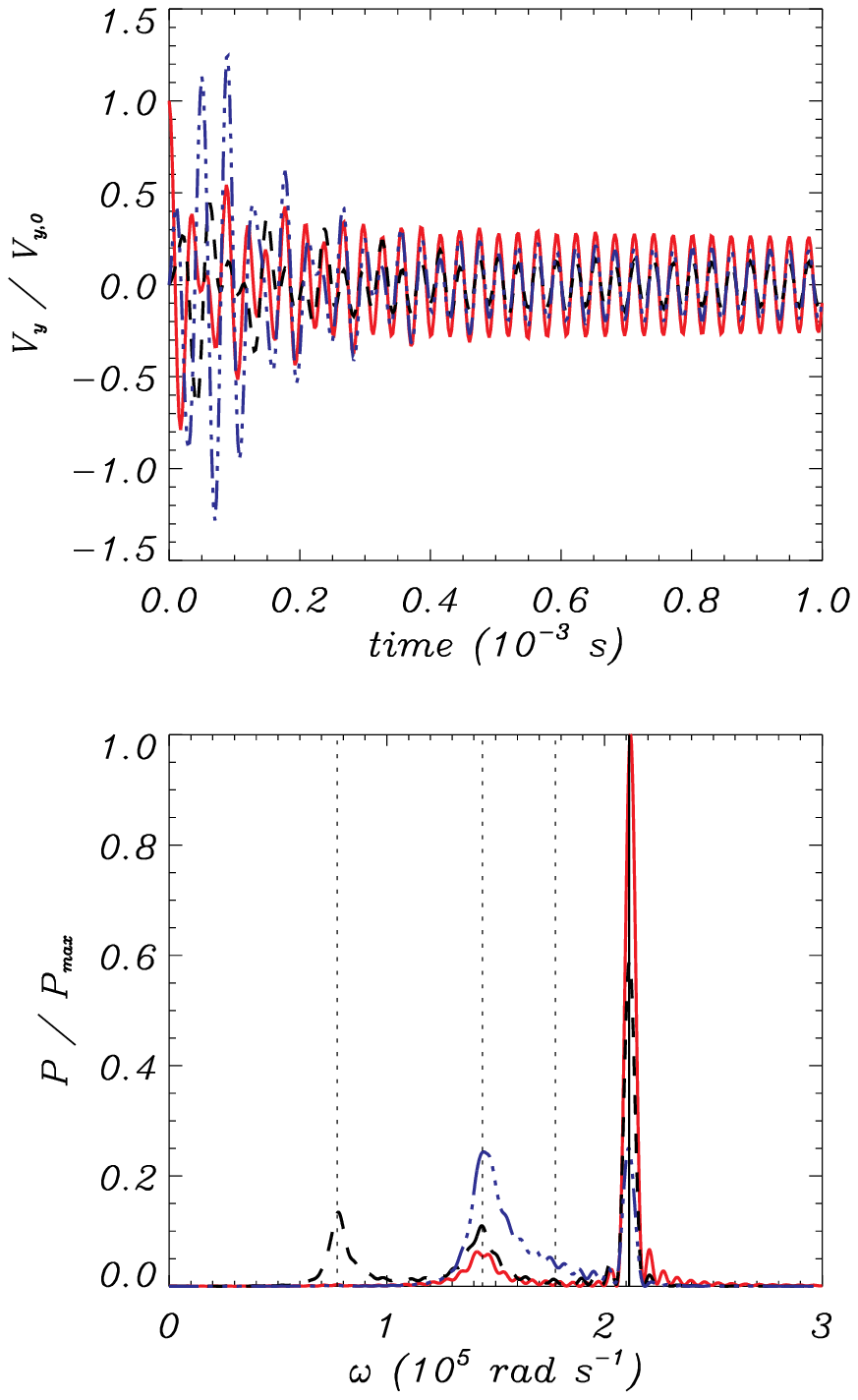}
		\caption{Results of a simulation of a three-ion plasma with upper chromospheric conditions. The top panels show the $y$-component of the velocity of ions and the bottom panels show the power spectra of the oscillations. On the left panels, the effect of collisions is ignored. On the right panels collisions are taken into account. The wavenumber is $k_{x}=\pi/5 \ \Rm{m^{-1}}$. The red solid, blue dot-dashed and black dashed lines correspond to $p$, $\Rm{He} \ \textsc{ii}$ and $\Rm{He} \ \textsc{iii}$, respectively.}
		\label{fig:sim_3ions}
	\end{figure*}

	Figure \ref{fig:sim_3ions} displays the results of two simulations with the initial perturbation
	
	\begin{equation} \label{eq:init_cond_3ions}
		V_{p,y}(t=0)=10^{-3}c_{\Rm{A}}\cos \left(k_{x}x\right),
	\end{equation}
	while the $\Rm{He} \ \textsc{ii}$ and the $\Rm{He} \ \textsc{iii}$ fluids are initially at rest. The chosen wave number is $k_{x}=\pi/5 \ \Rm{m^{-1}}$ (i.e., $\omega_{\Rm{A}}/\Omega_{p}\approx 0.46$, a regime where the results from the multi-fluid description already show a clear departure from ideal MHD). The left panels represent the case without collisions between the ions. The right panels show the results of the simulation in which the collisions are taken into account. The vertical lines on the bottom panels show the roots of the dispersion relation. The power spectrum on the left exhibits the expected four peaks, that agree with the solutions from the dispersion relations. On the right panels we see that the powers of the three L modes are much lower than on the left: they are strongly damped by cause of collisions.

	If we choose a much lower wavenumber, we enter the range at which the multi-fluid and the single-fluid approaches yield approximately the same results, with the exception of the extremely short relaxation time present in the former. Hence, the results of the simulations are quite similar to those displayed in Figures \ref{fig:sim_2ions_a} and \ref{fig:sim_2ions_b} for the case of a two-ion plasma. For this reason we do not show new figures to illustrate this case. After the relaxation time, the magnetic field produces a strong coupling on the three fluids so they behave as a single one. Then, their motion is well described by Equation (\ref{eq:vs_polarized2}) (or the related ones, (\ref{eq:vs_polarized3}) and (\ref{eq:fit})), with a velocity amplitude given by Equation (\ref{eq:wave_amp}). The main difference, apart from an evident new value of the Alfvén frequency, is that during the relaxation time it is possible to detect two oscillation modes, whose frequencies are approximately given by Equations (\ref{eq:weighted_1}) and (\ref{eq:weighted_2}), instead of the only one that is present when only two ions are considered. Due to the high collision frequencies, the damping times of those modes are tiny compared to the period of the Alfvén wave: for instance, if the wavenumber is $k_{x}=\pi/10^{5} \ \Rm{m^{-1}}$, the damping times are on the order of $\tau=4 \times 10^{-5} \ \Rm{s}$, while the period of the Alfvén wave is $P=2\pi/\omega \approx 0.82 \ \Rm{s}$.

\subsection{Periodic driver} \label{sec:sims_periodic}
	Here, we use a domain given by $x \in [0,L]$, and the effect of the driver is simulated by imposing at $x=0$ a perturbation that is a periodic function of time.

\begin{figure*}
		\includegraphics[width=\hsize]{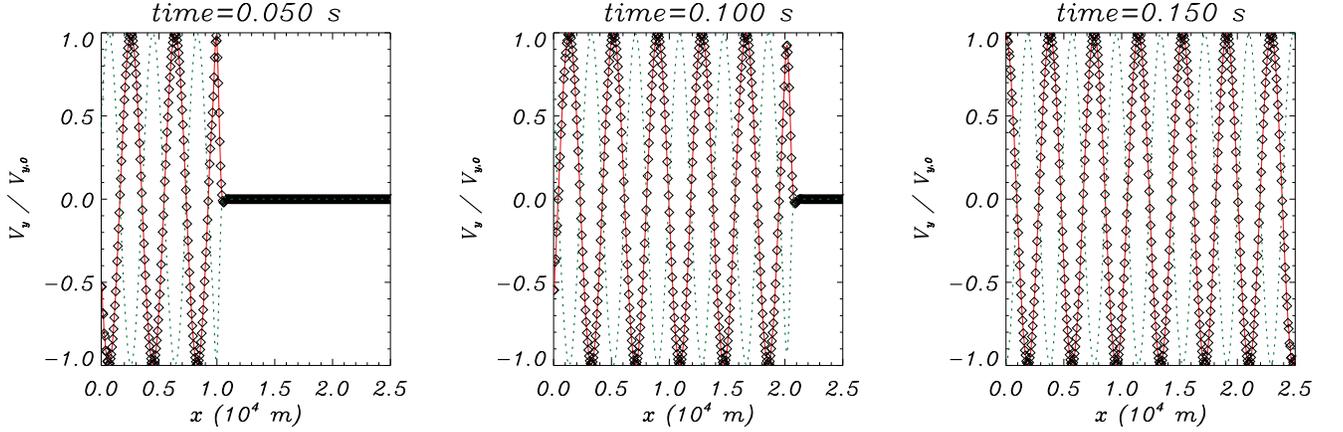}
		\caption{Wave generated by a periodic driver in a two-ion plasma with chromospheric conditions. The frequency of the driver is $\omega=10^{-3}\Omega_{p}$. The $y$-component of the velocity of the ions is shown as a function of the coordinate $x$. The red line represents the velocity of protons and the black symbols the velocity of the $\Rm{He} \ \textsc{ii}$ fluid. The green dotted lines show the normalized magnetic field perturbation, $B_{1,y}c_{\Rm{A}}/(V_{y,0}B_{x})$.}
		\label{fig:sim_driver_2ions_a}
	\end{figure*}

	We show in Figure \ref{fig:sim_driver_2ions_a} the results of a simulation with the parameters for the upper chromospheric region considering that the plasma is composed of only two species. The driver is given by
	\begin{equation} \label{eq:driver1}
		V_{p,y}(x=0,t)=V_{\Rm{He}\textsc{ii},y}(x=0,t)=10^{-3}c_{\Rm{A}} \cos \left(\omega t\right),
	\end{equation}
	with a frequency $\omega=10^{-3} \Omega_{p}$. The length of the domain is $L=2.5 \times 10^4 \ \Rm{m}$ and we use a uniform grid of $N=401$ points.

	Each frame of Figure \ref{fig:sim_driver_2ions_a} corresponds to a different time of the simulation. Comparing the positions of the wavefront (or the rightmost maximum of amplitude) we can check that the perturbation propagates at a phase speed of about $200 \ \Rm{km \ s^{-1}}$, which agrees well with the Alfvén speed of this plasma. The two fluids, protons and singly ionized helium, oscillate with the same phase and amplitude, in anti-phase with respect to the magnetic field perturbation. We fit the oscillation to a function $f(x)\sim \cos \left(k_{x}x\right)$ and obtain the wavenumber $k_{\Rm{x}}\approx 0.001643 \ \Rm{m^{-1}}$. From the dispersion relation we get two modes with $k_{R,+}=0.001644 \ \Rm{m^{-1}}$ and $k_{R,-}=0.001642 \ \Rm{m^{-1}}$, respectively. Thus, we check that the simulation is consistent with the solutions provided by the dispersion relation.

	As in the case of the impulsive driver, the resulting wave is a combination of the left-hand and the right-hand polarized modes, which yields the composition of a carrier wave and an envelope wave. Here, the wavenumber of the carrier wave is $\kappa_{C}=\left(k_{R,+}+k_{R,-}\right)/2$ and the wavenumber of the envelope wave is $\kappa_{E}=\left(k_{R,+}-k_{R,-}\right)/2$. The frequency chosen for the previous simulation leads to $\kappa_{E} \approx 1.49 \times 10^{-6} \ \Rm{m^{-1}}$, which is equivalent to a wavelength $\lambda_{E} \approx 4.2 \times 10^{6} \ \Rm{m}$. Such value is much larger than the length of the domain, and the existence of the envelope wave cannot be discerned in this simulation. Nonetheless, we have checked that it can be found in simulations with higher frequencies, at which there are greater differences between the wavenumbers of the two opposite polarized modes and, hence, $\lambda_{E}$ is much shorter.

	\begin{figure}
			\centering
			\includegraphics[width=0.5\hsize]{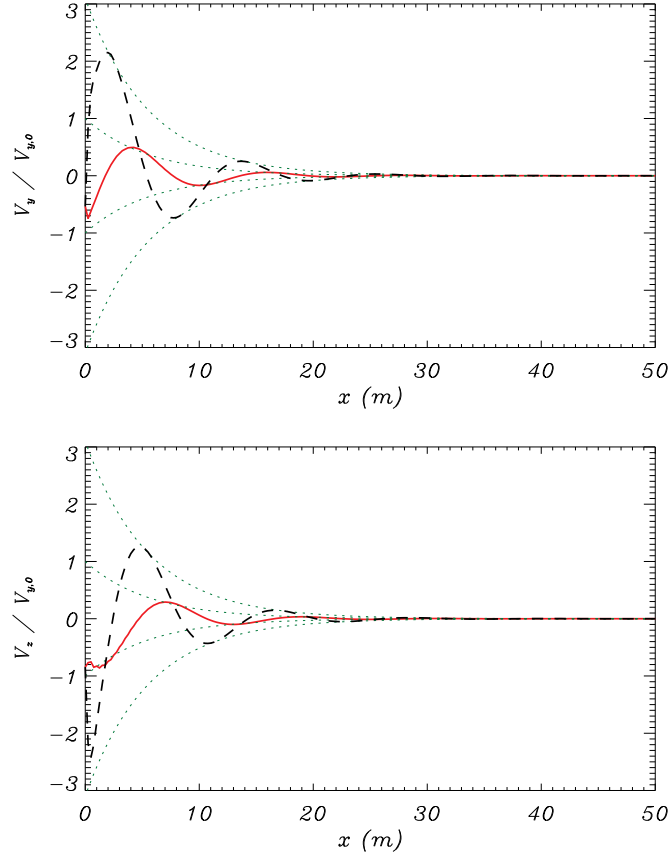}
		\caption{Velocities of ions at a time $t=10^{-3} \ \Rm{s}$ in a simulation in which the L-mode is excited by a driver with frequency $\omega=\Omega_{\Rm{He}\textsc{ii}}$. The $y$-components and $z$-component are shown in the top and the bottom panels, respectively. The red solid lines correspond to the protons and the black dashed lines correspond to the singly ionized helium. The dotted curves outline the spatial exponential decay computed through the dispersion relation, Equation (\ref{eq:dr_2ions_a_kx}). A movie with the full simulation is available in the web version of the paper.}
		\label{fig:sim_driver_L}
	\end{figure}

	\begin{figure}
		\centering
		\includegraphics[width=0.5\hsize]{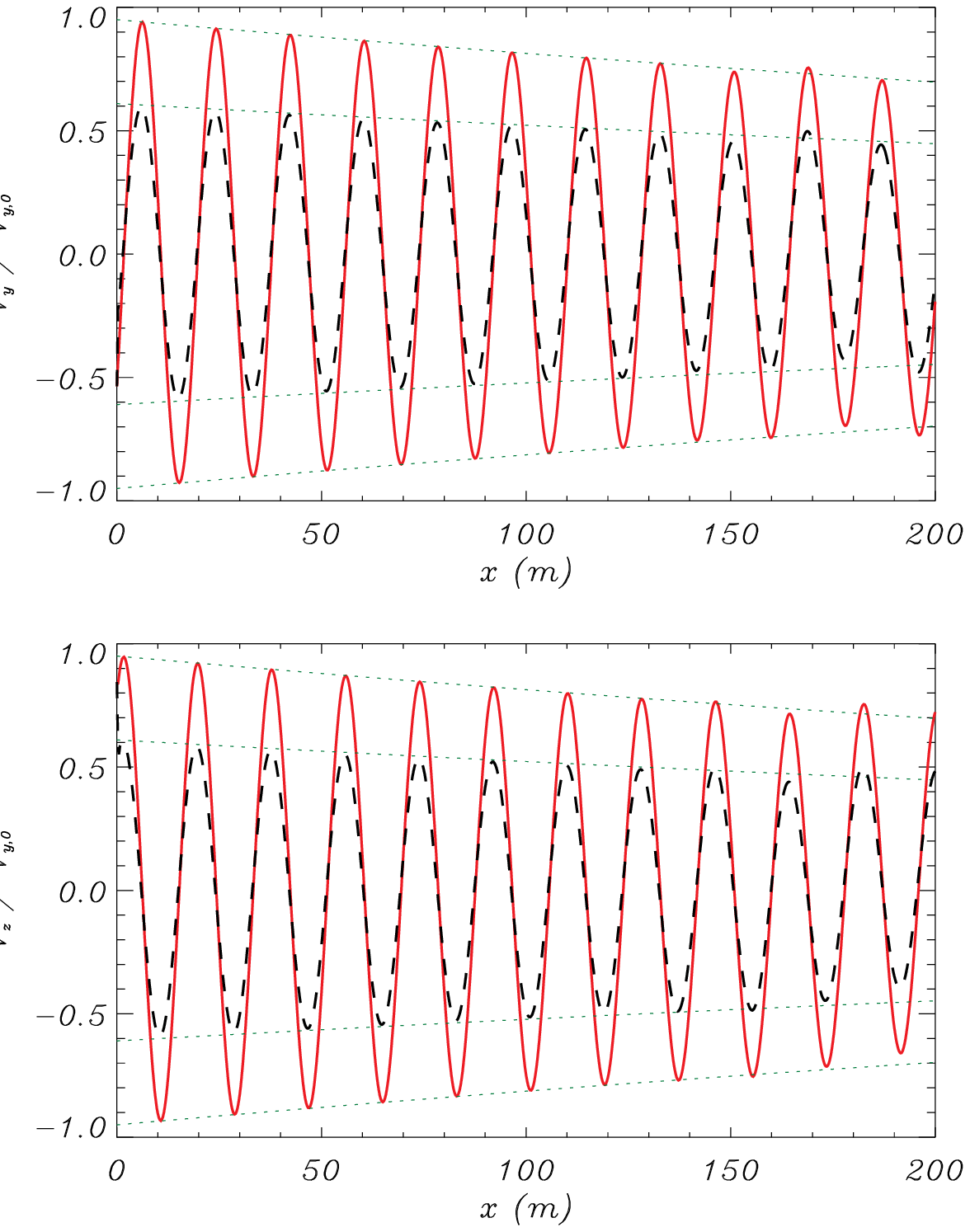}
		\caption{Velocities of ions at a time $t=10^{-3} \ \Rm{s}$ in a simulation of the R-mode at the frequency of resonance $\omega=\Omega_{\Rm{He}\textsc{ii}}$. The same color and line style as in Figure \ref{fig:sim_driver_L} is used here. This figure is available as an animation in the web version of this paper.}
		\label{fig:sim_driver_R}
	\end{figure}

	The most interesting values of frequency in the study of waves excited by a periodic driver are those where resonances appear in the collisionless case, i.e., when the frequency of the driver coincides with one of the cyclotron frequencies of the system. At these points, the behaviors of the two modes exhibit great dissimilarities, and it is convenient to analyze them separately. The driver employed in the previous simulations causes the excitation of both modes. So, it is not useful for this task. Nevertheless, it is possible to find other kind of drivers that allows to excite the chosen mode exclusively. For instance, to study only the left-hand mode (+) or only the right-hand mode (-), we may use the following configuration,
	\begin{equation} \label{eq:driver_V}
		\bm{V_{s,\pm}}(x=0,t)=\left( \begin{array}{c}
		0 \\
		V_{0} \cos \left(\omega t\right) \\
		\mp V_{0} \sin \left(\omega t\right)
		\end{array} \right)
	\end{equation}
	and
	\begin{equation} \label{eq:driver_B}
		\bm{B_{1,\pm}}(x=0,t)=\left( \begin{array}{c}
		0 \\
		B_{1,0} \cos \left(\omega t\right) \\
		\mp B_{1,0} \sin \left(\omega t\right)
		\end{array} \right),
	\end{equation}
	where the amplitudes of the perturbations are linked by $B_{1,0}=-B_{x}V_{0}/c_{\Rm{A}}$.

	In Figure \ref{fig:sim_driver_L} we show the results of a simulation of the L mode at the frequency of resonance $\omega=\Omega_{\Rm{He}\textsc{ii}}$. The effect of collisions is included and we use a total of $N=2001$ points to cover a domain of $L=500 \ \Rm{m}$, although only the interval $x \in[0,50]$ is represented in the plots; the motivation for using a bigger domain in the simulation than the one shown in the figure is to avoid the interference of unwanted numerical effects caused by the rightmost boundary. It can be checked that the perturbation is strongly damped in space: it does not propagate beyond a distance $x\approx 30 \ \Rm{m}$ from the origin. The dotted curves represent the exponential decay given by the dispersion relation and we see that it fits well the damped oscillation of the simulation. In addition, using those curves as a reference we can compute the amplitude ratio and check that it has a value $|V_{p}/V_{\Rm{He}\textsc{ii}}| \approx 0.35$, which agrees with the results displayed in Figure \ref{fig:amplitudes_periodic}. We also see the predicted phase shifts between the velocities of the two fluids. From Section \ref{sec:periodic_2ions} we know that when $\nu_{p\Rm{He}\textsc{ii}}=0$, the ion cyclotron wave does not propagate at all: its wavenumber tends to infinity and, therefore, its phase speed is zero.

	Figure \ref{fig:sim_driver_R} shows a simulation of the R mode at the resonance frequency. The differences with the previous figure are evident. The phase speed of this mode is higher and it propagates to much farther distances since the spatial damping is much lower. Another fact that contrast with the ion cyclotron mode is that here the protons oscillate with a larger amplitude than the singly ionized helium: the amplitude ratio is $|V_{p}/V_{\Rm{He}\textsc{ii}}| \approx 1.6$, again in good agreement with the analysis from the dispersion relation. The two fluids oscillate in phase.

	The behavior of the two waves at the upper resonance, $\omega=\Omega_{p}$, is analogous to what has been explained in the two preceding paragraphs. Thus, we do not include new figures to explain this case. It can be described by inspecting Figures \ref{fig:sim_driver_L} and \ref{fig:sim_driver_R} and taking into account that the wavenumbers would be larger, the damping lengths would be shorter and the amplitude ratio of the L mode would be now greater than unity.

	Finally, we briefly turn our attention to the case of three-ion plasmas. Contrary to what happens in the case of an impulsive driver, here the consideration of an additional ionized species to the system does not cause the appearance of an extra oscillation mode. Thus, the motions of all the species in the plasma are still governed by the combination of only one left-hand mode and one right-hand mode. The main difference is that a third frequency of resonance exists.

\begin{figure}
		\centering
		\includegraphics[width=0.5\hsize]{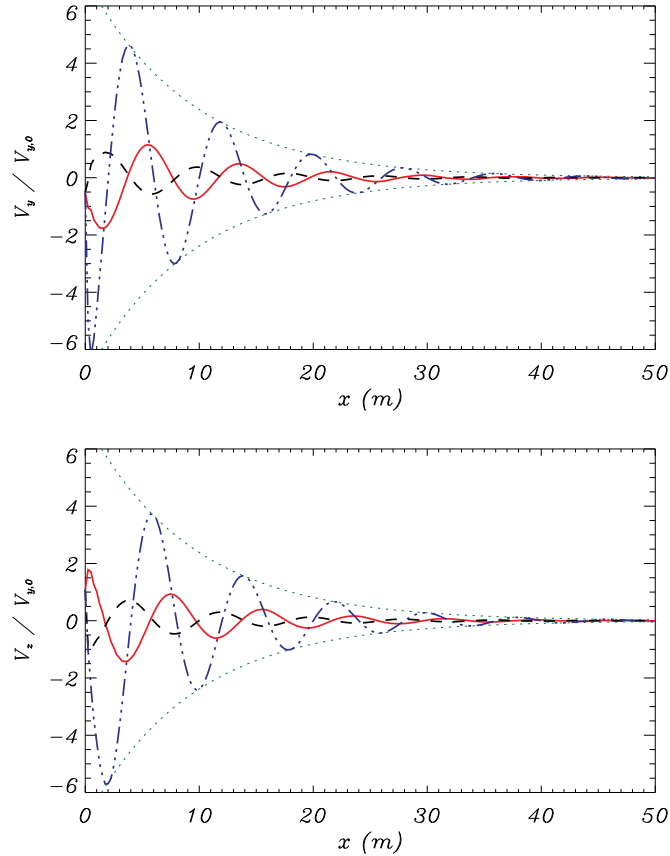}
		\caption{Velocities of ions at the time $t=10^{-3} \ \Rm{s}$ of a simulation of the L-mode at the frequency of resonance $\omega=\Omega_{\Rm{He}\textsc{iii}}$ in a three-ion plasma with chromospheric conditions. The red solid lines represent the $y$-component of the normalized velocity of the protons; the black dashed lines correspond to the velocity of singly ionized helium and the blue dotted-dashed lines correspond to the velocity of the doubly ionized helium. The dotted curves outline the spatial damping computed through Equation (\ref{eq:dispersion}).}
		\label{fig:sim_driver_3ions}
	\end{figure}

	Figure \ref{fig:sim_driver_3ions} shows a simulation of a left-hand polarized wave in a three-ion plasma with chromospheric conditions excited by a driver with a frequency $\omega=\Omega_{\Rm{He}\textsc{iii}}$. The driver is described by Equations (\ref{eq:driver_V}) and (\ref{eq:driver_B}) and we use $N=2001$ points to represent the domain $x \in [0,500]$ (again, the figure only shows a fraction of such domain). As in the simulation for $\omega=\Omega_{\Rm{He}\textsc{ii}}$ in the two-ions case (Figure \ref{fig:sim_driver_L}), we find that there is a strong spatial damping and the perturbation cannot propagate far from where it has been originated. Here, the ion that oscillates with a larger amplitude is the doubly ionized helium, as it would be expected since the driver is exciting a wave with the cyclotron frequency of that species.

\section{Conclusions} \label{sec:concl}
	This paper is the first of a series devoted to the comprehensive study of multi-fluid effects on the behavior and properties of high-frequency waves in plasmas. In this initial work, the multi-fluid approach has been applied to the investigation of incompressible waves in a fully ionized plasma. The subject of multi-ion plasmas has already been studied by numerous authors but most of them have focused exclusively on low frequency Alfvén waves \citep[e.g.,][]{1982JGR....87.5023I}, have analyzed plasmas composed of only two different ionized species \citep[e.g.,][]{2010Rahbarnia} or have not included the effect of collisions between the ions \citep[e.g.,][]{1973Ap&SS..20..391W,2001paw..book.....C}. On the contrary, we have explored a wide range of frequencies that goes from the low frequency Alfvén waves to the high frequency ion cyclotron and whistler waves, and our model takes into consideration the collisional interactions between the distinct species. Hence, it is more general than the approaches commonly employed in previous works and it can be applicable to a great number of astrophysical and laboratory plasmas. However, we have directed our attention to plasmas in the solar atmosphere, e.g., the upper chromospheric region, the lower corona and the solar wind at 1 AU.
	
	We have presented the system of non-linear equations that governs the temporal evolution of each species in a multi-ion plasma. From them, we have derived the dispersion relations that characterize the propagation of small-amplitude incompressible perturbations along the direction of the background magnetic field in an homogeneous medium. Those dispersion relations have allowed us to analyze the properties of waves excited both by an impulsive driver and by a periodic driver. In addition, we have compared the results given by those formulas with the predictions provided by the single-fluid model of ideal MHD. The first difference between the two approaches is that ideal MHD predicts the existence of the same number of oscillation modes independently of the kind of driver chosen, and it is not necessary to study the two cases separately as they are equivalent. This symmetry is not conserved in the multi-fluid model. The number of modes in the case of the periodic driver is independent of how many species compose the plasma. But when the effects of an impulsive driver are investigated, the number of solutions given by the dispersion relations increases with each additional ionized species considered. Moreover, from ideal MHD we obtain waves that are linearly polarized, while the waves from the multi-fluid description are circularly polarized and they show a clearly different behavior depending on whether their polarization is left-handed or right-handed.
	
	Apart from the analysis of the various dispersion relations, we have also performed numerical simulations to compute the temporal evolution of the perturbations in the plasma. The MolMHD code, initially developed to solve the ideal MHD equations \citep{Bona20092266}, has been extended to account for multi-fluid effects. In the present work, we have tested the results from the numerical code against those from the analytic dispersion relations. We checked that the oscillation frequencies and the amplitude ratios of the waves appearing in the simulations are consistent with the solutions given by the dispersion relations. If exposed to perturbations with frequencies much lower than the cyclotron frequencies, the various ionized species react as if they were a single fluid since they are strongly coupled by means of the magnetic field: they oscillate at the Alfvén frequency, in phase and with the same velocity amplitude, which is given by Equation (\ref{eq:wave_amp}). At higher frequencies, the interaction through the magnetic field is not enough to keep all the fluids as tightly coupled as before and the amplitude and the phase of the oscillations are different for each species and for each mode. At this regime the effect of elastic collisions should not be neglected since the frictional force associated to the velocity drifts may lead to a intense damping of the waves. We have found that this damping is stronger for the modes with the left-hand polarization than for the right-hand modes.
	
	We have computed the friction coefficient between the ions in the three solar plasmas of interest for this work. Due to the small value of this parameter in the lower solar corona, we have found that the effect of collisions is relevant only for times longer than several periods of the Alfvén wave. The friction coefficient in the solar wind at 1 AU is even smaller, so this environment can be treated as a collisionless fluid since the damping times of the oscillations are on the order of $10^{6} \ \Rm{s}$ or larger. Nevertheless, the multi-fluid model is generally still required to illustrate the properties of waves in those plasmas, in view of the fact that only perturbations with wavelengths larger than the critical value given by Equation (\ref{eq:lambda_min}) are described with a reasonable accuracy by ideal MHD. On the contrary, the collision frequencies computed using the parameters corresponding to an upper chromospheric region are not negligible in comparison with the cyclotron frequencies and, consequently, friction has a strong impact in the properties of the oscillation modes.
	
	There is another reason to take into account the effect of elastic collisions: when a periodic driver is considered, the momentum transfer removes the resonances and the strict cut-offs that appear in the collisionless case. Due to the diffusive effect of collisions, left-hand waves generated by a driver with a frequency that coincides with any of the cyclotron frequencies or is in the range of the cut-offs can propagate, instead of having a null phase speed or being evanescent. However, they are still strongly damped in space.
	
	The model developed and employed in this work provides more accurate results than the ones used in previous investigations. But, since we have also resorted to some simplifications, it still can be improved along several lines. For instance, we have not taken into account some effects produced by other kinds of collisions, like resistivity, caused by the interaction between ions and electrons and that is expected to be of importance for even higher frequencies than the ones explored here, or viscosity, that is due to self-collisions. Furthermore, we have focused exclusively on small-amplitude perturbations on fully ionized plasmas, while partial ionization effects have been shown to be of great relevance even for low frequency waves \citep[see, e.g.,][]{1956MNRAS.116..314P,2009ApJ...699.1553S,2011A&A...534A..93Z,2013ApJS..209...16S,2013ApJ...767..171S} and the non-linear regime is also of great interest, specially in respect to the matter of the heating of the plasma. In forthcoming papers we will address those issues.
	
	We remind also that we have only studied the case of a homogeneous medium but that it would be possible to apply our code to inhomogeneous media. In the presence of  inhomogeneities, that can be caused, for example, by gravitational stratification, effects like wave amplification (e.g., amplitude of waves propagating upwards in the solar corona is amplified due to the decrease in density with height), shock generation, refraction and reflection of waves or parametric decay may appear. Some of these effects have been investigated through a single-fluid MHD approach \citep{2005ApJ...632L..49S,2006JGRA..111.6101S} but it may be expected that inhomogeneities have a different impact on each ionized species of the plasma in the range of frequencies where the multi-fluid model is more appropriate than the single-fluid approximation. Take the solar wind as an example of inhomogeneous non-static medium: \citet{2007JGRA..112.8102H} performed a numerical study of the reflection of Alfvén waves and found that outward propagating waves become less dominant than sunward propagating waves for distances from the Sun beyond the Alfvén critical point (where the plasma flow velocity is equal to the Alfvén speed), in agreement with the observations; in addition, \citet{2007ApJ...661.1222L} investigated the effect of the differential ion flow using a two-fluid model and found that at large distances beyond the Alfvénic point low frequency waves tend to equalize the speeds of the ions. It would be of interest the application of the multi-fluid model to the investigation of the effects of reflection and differential ion flow on high frequency waves.

\acknowledgements
	The authors are grateful to the anonymous referee for the comprehensive and insightful remarks, which have helped us to improve the present work. We acknowledge the support from MINECO and FEDER funds through grant AYA2014-54485-P. D.M. acknowledges support from MINECO through a “FPI” grant. R.S. acknowledges support from MINECO and UIB through a ``Ramón y Cajal" grant (RYC-2014-14970). J.T. acknowledges support from MINECO and UIB through a "Ramón y Cajal" grant.
	
%\bibpunct{(}{)}{;}{a}{}{,} %to follow the A&A bib style
\bibliographystyle{aasjournal}
\bibliography{mybib2}

\begin{thebibliography}{}
\expandafter\ifx\csname natexlab\endcsname\relax\def\natexlab#1{#1}\fi

\bibitem[{{Abbo} {et~al.}(2016){Abbo}, {Ofman}, {Antiochos}, {Hansteen},
  {Harra}, {Ko}, {Lapenta}, {Li}, {Riley}, {Strachan}, {von Steiger}, \&
  {Wang}}]{2016SSRv..tmp...34A}
{Abbo}, L., {Ofman}, L., {Antiochos}, S.~K., {et~al.} 2016, \ssr,
  doi:10.1007/s11214-016-0264-1

\bibitem[{{Aellig} {et~al.}(2001){Aellig}, {Lazarus}, \&
  {Steinberg}}]{2001GeoRL..28.2767A}
{Aellig}, M.~R., {Lazarus}, A.~J., \& {Steinberg}, J.~T. 2001, \grl, 28, 2767

\bibitem[{{Ahmad}(1977)}]{1977SoPh...53..409A}
{Ahmad}, I.~A. 1977, \solphys, 53, 409

\bibitem[{{Alfv{\'e}n}(1942)}]{1942Natur.150..405A}
{Alfv{\'e}n}, H. 1942, \nat, 150, 405

\bibitem[{{Anders} \& {Grevesse}(1989)}]{1989GeCoA..53..197A}
{Anders}, E., \& {Grevesse}, N. 1989, \gca, 53, 197

\bibitem[{{Arons} \& {Max}(1975)}]{1975ApJ...196L..77A}
{Arons}, J., \& {Max}, C.~E. 1975, Astrophysical Journal Letters, 196, L77

\bibitem[{{Axford} \& {McKenzie}(1992)}]{1992sws..coll....1A}
{Axford}, W.~I., \& {McKenzie}, J.~F. 1992, in Solar Wind Seven Colloquium, ed.
  E.~{Marsch} \& R.~{Schwenn}, 1--5

\bibitem[{{Balsara}(1996)}]{1996ApJ...465..775B}
{Balsara}, D.~S. 1996, \apj, 465, 775

\bibitem[{{Barakat} \& {Schunk}(1982)}]{1982PlPh...24..389B}
{Barakat}, A.~R., \& {Schunk}, R.~W. 1982, Plasma Physics, 24, 389

\bibitem[{{Belcher} \& {Davis}(1971)}]{1971JGR....76.3534B}
{Belcher}, J.~W., \& {Davis}, Jr., L. 1971, \jgr, 76, 3534

\bibitem[{{Berthold} {et~al.}(1960){Berthold}, {Harris}, \&
  {Hope}}]{1960JGR....65.2233B}
{Berthold}, W.~K., {Harris}, A.~K., \& {Hope}, H.~J. 1960, \jgr, 65, 2233

\bibitem[{Bona {et~al.}(2009)Bona, Bona-Casas, \& Terradas}]{Bona20092266}
Bona, C., Bona-Casas, C., \& Terradas, J. 2009, Journal of Computational
  Physics, 228, 2266

\bibitem[{Bostick \& Levine(1952)}]{1952PhysRev.87.671}
Bostick, W.~H., \& Levine, M.~A. 1952, Phys. Rev., 87, 671

\bibitem[{{Braginskii}(1965)}]{1965RvPP....1..205B}
{Braginskii}, S.~I. 1965, Reviews of Plasma Physics, 1, 205

\bibitem[{Callen(2006)}]{Callen2006a}
Callen, J.~D. 2006, in Fundamentals of Plasma Physics, draft edn. (Madison,
  Wisconsin)

\bibitem[{{Chen} \& {Hasegawa}(1974)}]{1974PhFl...17.1399C}
{Chen}, L., \& {Hasegawa}, A. 1974, Physics of Fluids, 17, 1399

\bibitem[{{Chen} \& {Hu}(2001)}]{2001SoPh..199..371C}
{Chen}, Y., \& {Hu}, Y.~Q. 2001, \solphys, 199, 371

\bibitem[{{Chmyrev} {et~al.}(1988){Chmyrev}, {Bilichenko}, {Pokhotelov},
  {Marchenko}, \& {Lazarev}}]{1988PhyS...38..841C}
{Chmyrev}, V.~M., {Bilichenko}, S.~V., {Pokhotelov}, V.~I., {Marchenko}, V.~A.,
  \& {Lazarev}, V.~I. 1988, \physscr, 38, 841

\bibitem[{Coleman(1966)}]{1966PhysRevLett.17.207}
Coleman, P.~J. 1966, Phys. Rev. Lett., 17, 207

\bibitem[{{Courant} {et~al.}(1928){Courant}, {Friedrichs}, \&
  {Lewy}}]{1928MatAn.100...32C}
{Courant}, R., {Friedrichs}, K., \& {Lewy}, H. 1928, Mathematische Annalen,
  100, 32

\bibitem[{{Cramer}(2001)}]{2001paw..book.....C}
{Cramer}, N.~F. 2001, {The Physics of Alfv{\'e}n Waves}, 312

\bibitem[{{De Pontieu} {et~al.}(2001){De Pontieu}, {Martens}, \&
  {Hudson}}]{2001ApJ...558..859D}
{De Pontieu}, B., {Martens}, P.~C.~H., \& {Hudson}, H.~S. 2001, \apj, 558, 859

\bibitem[{{De Pontieu} {et~al.}(2007){De Pontieu}, {McIntosh}, {Carlsson},
  {Hansteen}, {Tarbell}, {Schrijver}, {Title}, {Shine}, {Tsuneta}, {Katsukawa},
  {Ichimoto}, {Suematsu}, {Shimizu}, \& {Nagata}}]{2007Sci...318.1574D}
{De Pontieu}, B., {McIntosh}, S.~W., {Carlsson}, M., {et~al.} 2007, Science,
  318, 1574

\bibitem[{{Demars} \& {Schunk}(1979)}]{1979JPhD...12.1051D}
{Demars}, H.~G., \& {Schunk}, R.~W. 1979, Journal of Physics D Applied Physics,
  12, 1051

\bibitem[{{Demars} \& {Schunk}(1994)}]{1994JGR....99.2215D}
---. 1994, \jgr, 99, 2215

\bibitem[{{Draine}(1986)}]{1986MNRAS.220..133D}
{Draine}, B.~T. 1986, \mnras, 220, 133

\bibitem[{{Dzhalilov} {et~al.}(2008){Dzhalilov}, {Kuznetsov}, \&
  {Staude}}]{2008A&A...489..769D}
{Dzhalilov}, N.~S., {Kuznetsov}, V.~D., \& {Staude}, J. 2008, \aap, 489, 769

\bibitem[{{Echim} {et~al.}(2011){Echim}, {Lemaire}, \&
  {Lie-Svendsen}}]{2011SGeo...32....1E}
{Echim}, M.~M., {Lemaire}, J., \& {Lie-Svendsen}, {\O}. 2011, Surveys in
  Geophysics, 32, 1

\bibitem[{{Fludra} {et~al.}(1999){Fludra}, {Del Zanna}, {Alexander}, \&
  {Bromage}}]{1999JGR...104.9709F}
{Fludra}, A., {Del Zanna}, G., {Alexander}, D., \& {Bromage}, B.~J.~I. 1999,
  \jgr, 104, 9709

\bibitem[{{Fontenla} {et~al.}(1993){Fontenla}, {Avrett}, \&
  {Loeser}}]{1993ApJ...406..319F}
{Fontenla}, J.~M., {Avrett}, E.~H., \& {Loeser}, R. 1993, \apj, 406, 319

\bibitem[{{Forteza} {et~al.}(2007){Forteza}, {Oliver}, {Ballester}, \&
  {Khodachenko}}]{2007A&A...461..731F}
{Forteza}, P., {Oliver}, R., {Ballester}, J.~L., \& {Khodachenko}, M.~L. 2007,
  \aap, 461, 731

\bibitem[{{Ganguli}(1996)}]{1996RvGeo..34..311G}
{Ganguli}, S.~B. 1996, Reviews of Geophysics, 34, 311

\bibitem[{{Ganguli} \& {Palmadesso}(1988)}]{1988AdSpR...8...69G}
{Ganguli}, S.~B., \& {Palmadesso}, P.~J. 1988, Advances in Space Research, 8,
  69

\bibitem[{Gekelman(1999)}]{1999JGRA:JGRA14819}
Gekelman, W. 1999, Journal of Geophysical Research: Space Physics, 104, 14417

\bibitem[{{Goedbloed} \& {Poedts}(2004)}]{2004prma.book.....G}
{Goedbloed}, J.~P.~H., \& {Poedts}, S. 2004, {Principles of
  Magnetohydrodynamics}

\bibitem[{{Goncalves} {et~al.}(1993){Goncalves}, {Jatenco-Pereira}, \&
  {Opher}}]{1993A&A...279..351G}
{Goncalves}, D.~R., {Jatenco-Pereira}, V., \& {Opher}, R. 1993, \aap, 279, 351

\bibitem[{{Goossens} {et~al.}(2012){Goossens}, {Andries}, {Soler}, {Van
  Doorsselaere}, {Arregui}, \& {Terradas}}]{2012ApJ...753..111G}
{Goossens}, M., {Andries}, J., {Soler}, R., {et~al.} 2012, \apj, 753, 111

\bibitem[{{Gurnett} \& {Goertz}(1981)}]{1981JGR....86..717G}
{Gurnett}, D.~A., \& {Goertz}, C.~K. 1981, \jgr, 86, 717

\bibitem[{Harten(1983)}]{Harten:1983:HRS}
Harten, A. 1983, 49, 357

\bibitem[{{Hartle} \& {Sturrock}(1968)}]{1968ApJ...151.1155H}
{Hartle}, R.~E., \& {Sturrock}, P.~A. 1968, \apj, 151, 1155

\bibitem[{{Heidbrink}(2008)}]{2008PhPl...15e5501H}
{Heidbrink}, W.~W. 2008, Physics of Plasmas, 15, 055501

\bibitem[{{Hollweg} \& {Isenberg}(2002)}]{2002JGRA..107.1147H}
{Hollweg}, J.~V., \& {Isenberg}, P.~A. 2002, Journal of Geophysical Research
  (Space Physics), 107, 1147

\bibitem[{{Hollweg} \& {Isenberg}(2007)}]{2007JGRA..112.8102H}
---. 2007, Journal of Geophysical Research (Space Physics), 112, A08102

\bibitem[{{Isenberg} \& {Hollweg}(1982)}]{1982JGR....87.5023I}
{Isenberg}, P.~A., \& {Hollweg}, J.~V. 1982, \jgr, 87, 5023

\bibitem[{{Isenberg} \& {Hollweg}(1983)}]{1983JGR....88.3923I}
---. 1983, \jgr, 88, 3923

\bibitem[{{Jatenco-Pereira} \& {Opher}(1989)}]{1989MNRAS.236....1J}
{Jatenco-Pereira}, V., \& {Opher}, R. 1989, \mnras, 236, 1

\bibitem[{{Jephcott}(1959)}]{1959Natur.183.1652J}
{Jephcott}, D.~F. 1959, \nat, 183, 1652

\bibitem[{{Jess} {et~al.}(2009){Jess}, {Mathioudakis}, {Erd{\'e}lyi},
  {Crockett}, {Keenan}, \& {Christian}}]{2009Sci...323.1582J}
{Jess}, D.~B., {Mathioudakis}, M., {Erd{\'e}lyi}, R., {et~al.} 2009, Science,
  323, 1582

\bibitem[{{Khomenko} {et~al.}(2014){Khomenko}, {Collados}, {D{\'{\i}}az}, \&
  {Vitas}}]{2014PhPl...21i2901K}
{Khomenko}, E., {Collados}, M., {D{\'{\i}}az}, A., \& {Vitas}, N. 2014, Physics
  of Plasmas, 21, 092901

\bibitem[{{Konikov} {et~al.}(1989){Konikov}, {Gorbachev}, {Khazanov}, \&
  {Chernov}}]{1989P&SS...37.1157K}
{Konikov}, I.~V., {Gorbachev}, O.~A., {Khazanov}, G.~V., \& {Chernov}, A.~A.
  1989, \planss, 37, 1157

\bibitem[{{Krti{\v c}ka} \& {Kub{\'a}t}(2000)}]{2000A&A...359..983K}
{Krti{\v c}ka}, J., \& {Kub{\'a}t}, J. 2000, \aap, 359, 983

\bibitem[{{Krti{\v c}ka} \& {Kub{\'a}t}(2001)}]{2001A&A...369..222K}
---. 2001, \aap, 369, 222

\bibitem[{{Kulsrud} \& {Pearce}(1969)}]{1969ApJ...156..445K}
{Kulsrud}, R., \& {Pearce}, W.~P. 1969, \apj, 156, 445

\bibitem[{{Laming} \& {Feldman}(2003)}]{2003ApJ...591.1257L}
{Laming}, J.~M., \& {Feldman}, U. 2003, \apj, 591, 1257

\bibitem[{{Leer} \& {Axford}(1972)}]{1972SoPh...23..238L}
{Leer}, E., \& {Axford}, W.~I. 1972, \solphys, 23, 238

\bibitem[{{Lehnert}(1954)}]{1954PhRv...94..815L}
{Lehnert}, B. 1954, Physical Review, 94, 815

\bibitem[{{Li} {et~al.}(2007){Li}, {Habbal}, \& {Li}}]{2007ApJ...661..593L}
{Li}, B., {Habbal}, S.~R., \& {Li}, X. 2007, \apj, 661, 593

\bibitem[{{Li} \& {Li}(2006)}]{2006A&A...456..359L}
{Li}, B., \& {Li}, X. 2006, \aap, 456, 359

\bibitem[{{Li} \& {Li}(2007)}]{2007ApJ...661.1222L}
---. 2007, \apj, 661, 1222

\bibitem[{{Li} \& {Li}(2008)}]{2008ApJ...682..667L}
---. 2008, \apj, 682, 667

\bibitem[{{Li} \& {Li}(2009)}]{2009A&A...494..361L}
---. 2009, \aap, 494, 361

\bibitem[{{Li} {et~al.}(2004){Li}, {Li}, {Hu}, \&
  {Habbal}}]{2004JGRA..109.7103L}
{Li}, B., {Li}, X., {Hu}, Y.-Q., \& {Habbal}, S.~R. 2004, Journal of
  Geophysical Research (Space Physics), 109, A07103

\bibitem[{{Li} {et~al.}(2006){Li}, {Li}, \& {Labrosse}}]{2006JGRA..111.8106L}
{Li}, B., {Li}, X., \& {Labrosse}, N. 2006, Journal of Geophysical Research
  (Space Physics), 111, A08106

\bibitem[{{Lighthill}(1960)}]{1960RSPTA.252..397L}
{Lighthill}, M.~J. 1960, Philosophical Transactions of the Royal Society of
  London Series A, 252, 397

\bibitem[{{Lundquist}(1949)}]{1949Natur.164..145L}
{Lundquist}, S. 1949, \nat, 164, 145

\bibitem[{{McIntosh} {et~al.}(2011){McIntosh}, {de Pontieu}, {Carlsson},
  {Hansteen}, {Boerner}, \& {Goossens}}]{2011Natur.475..477M}
{McIntosh}, S.~W., {de Pontieu}, B., {Carlsson}, M., {et~al.} 2011, \nat, 475,
  477

\bibitem[{{Meng} {et~al.}(2015){Meng}, {van der Holst}, {T{\'o}th}, \&
  {Gombosi}}]{2015MNRAS.454.3697M}
{Meng}, X., {van der Holst}, B., {T{\'o}th}, G., \& {Gombosi}, T.~I. 2015,
  \mnras, 454, 3697

\bibitem[{{Mouschovias} {et~al.}(2011){Mouschovias}, {Ciolek}, \&
  {Morton}}]{2011MNRAS.415.1751M}
{Mouschovias}, T.~C., {Ciolek}, G.~E., \& {Morton}, S.~A. 2011, \mnras, 415,
  1751

\bibitem[{{Ofman}(2004{\natexlab{a}})}]{2004AdSpR..33..681O}
{Ofman}, L. 2004{\natexlab{a}}, Advances in Space Research, 33, 681

\bibitem[{{Ofman}(2004{\natexlab{b}})}]{2004JGRA..109.7102O}
---. 2004{\natexlab{b}}, Journal of Geophysical Research (Space Physics), 109,
  A07102

\bibitem[{{Olsen} \& {Leer}(1999)}]{1999JGR...104.9963O}
{Olsen}, E.~L., \& {Leer}, E. 1999, \jgr, 104, 9963

\bibitem[{{Piddington}(1956)}]{1956MNRAS.116..314P}
{Piddington}, J.~H. 1956, \mnras, 116, 314

\bibitem[{{Pudritz}(1990)}]{1990ApJ...350..195P}
{Pudritz}, R.~E. 1990, \apj, 350, 195

\bibitem[{Rahbarnia {et~al.}(2010)Rahbarnia, Ullrich, Sauer, Grulke, \&
  Klinger}]{2010Rahbarnia}
Rahbarnia, K., Ullrich, S., Sauer, K., Grulke, O., \& Klinger, T. 2010, Physics
  of Plasmas, 17, doi:http://dx.doi.org/10.1063/1.3322852

\bibitem[{{Sarmin} \& {Chudov}(1963)}]{1963SarminChudov}
{Sarmin}, E.~N., \& {Chudov}, L.~A. 1963, U.S.S.R. Comput. Math. Math. Phys.,
  3, 1537

\bibitem[{Schiesser(1991)}]{schiesser1991numerical}
Schiesser, W. 1991, The Numerical Method of Lines: Integration of Partial
  Differential Equations (Academic Press)

\bibitem[{{Schunk}(1977)}]{1977RvGSP..15..429S}
{Schunk}, R.~W. 1977, Reviews of Geophysics and Space Physics, 15, 429

\bibitem[{{Sittler} \& {Guhathakurta}(1999)}]{1999ApJ...523..812S}
{Sittler}, Jr., E.~C., \& {Guhathakurta}, M. 1999, \apj, 523, 812

\bibitem[{{Soler} {et~al.}(2013{\natexlab{a}}){Soler}, {Carbonell}, \&
  {Ballester}}]{2013ApJS..209...16S}
{Soler}, R., {Carbonell}, M., \& {Ballester}, J.~L. 2013{\natexlab{a}}, \apjs,
  209, 16

\bibitem[{{Soler} {et~al.}(2013{\natexlab{b}}){Soler}, {Carbonell},
  {Ballester}, \& {Terradas}}]{2013ApJ...767..171S}
{Soler}, R., {Carbonell}, M., {Ballester}, J.~L., \& {Terradas}, J.
  2013{\natexlab{b}}, \apj, 767, 171

\bibitem[{{Soler} {et~al.}(2009){Soler}, {Oliver}, \&
  {Ballester}}]{2009ApJ...699.1553S}
{Soler}, R., {Oliver}, R., \& {Ballester}, J.~L. 2009, \apj, 699, 1553

\bibitem[{{Soler} {et~al.}(2010){Soler}, {Oliver}, \&
  {Ballester}}]{2010A&A...512A..28S}
---. 2010, \aap, 512, A28

\bibitem[{{Spitzer}(1962)}]{1962pfig.book.....S}
{Spitzer}, L. 1962, {Physics of Fully Ionized Gases}

\bibitem[{{Stix}(1992)}]{1992wapl.book.....S}
{Stix}, T.~H. 1992, {Waves in plasmas}

\bibitem[{{Sturrock} \& {Hartle}(1966)}]{1966PhRvL..16..628S}
{Sturrock}, P.~A., \& {Hartle}, R.~E. 1966, Physical Review Letters, 16, 628

\bibitem[{{Suzuki} \& {Inutsuka}(2005)}]{2005ApJ...632L..49S}
{Suzuki}, T.~K., \& {Inutsuka}, S.-i. 2005, Astrophysical Journal Letters, 632,
  L49

\bibitem[{{Suzuki} \& {Inutsuka}(2006)}]{2006JGRA..111.6101S}
{Suzuki}, T.~K., \& {Inutsuka}, S.-I. 2006, Journal of Geophysical Research
  (Space Physics), 111, A06101

\bibitem[{{Tomczyk} {et~al.}(2007){Tomczyk}, {McIntosh}, {Keil}, {Judge},
  {Schad}, {Seeley}, \& {Edmondson}}]{2007Sci...317.1192T}
{Tomczyk}, S., {McIntosh}, S.~W., {Keil}, S.~L., {et~al.} 2007, Science, 317,
  1192

\bibitem[{{Tu}(1987)}]{1987SoPh..109..149T}
{Tu}, C.-Y. 1987, \solphys, 109, 149

\bibitem[{{Tu} \& {Marsch}(1997)}]{1997SoPh..171..363T}
{Tu}, C.-Y., \& {Marsch}, E. 1997, \solphys, 171, 363

\bibitem[{{Underhill}(1983)}]{1983ApJ...268L.127U}
{Underhill}, A.~B. 1983, Astrophysical Journal Letters, 268, L127

\bibitem[{{Usmanov} {et~al.}(2000){Usmanov}, {Goldstein}, {Besser}, \&
  {Fritzer}}]{2000JGR...10512675U}
{Usmanov}, A.~V., {Goldstein}, M.~L., {Besser}, B.~P., \& {Fritzer}, J.~M.
  2000, \jgr, 105, 12675

\bibitem[{{van der Holst} {et~al.}(2014){van der Holst}, {Sokolov}, {Meng},
  {Jin}, {Manchester}, {T{\'o}th}, \& {Gombosi}}]{2014ApJ...782...81V}
{van der Holst}, B., {Sokolov}, I.~V., {Meng}, X., {et~al.} 2014, \apj, 782, 81

\bibitem[{{Van Doorsselaere} {et~al.}(2008){Van Doorsselaere}, {Nakariakov}, \&
  {Verwichte}}]{2008ApJ...676L..73V}
{Van Doorsselaere}, T., {Nakariakov}, V.~M., \& {Verwichte}, E. 2008,
  Astrophysical Journal Letters, 676, L73

\bibitem[{{V{\'a}squez} {et~al.}(2003){V{\'a}squez}, {van Ballegooijen}, \&
  {Raymond}}]{2003ApJ...598.1361V}
{V{\'a}squez}, A.~M., {van Ballegooijen}, A.~A., \& {Raymond}, J.~C. 2003,
  \apj, 598, 1361

\bibitem[{{Vranjes} \& {Krstic}(2013)}]{2013A&A...554A..22V}
{Vranjes}, J., \& {Krstic}, P.~S. 2013, \aap, 554, A22

\bibitem[{{Warmuth} \& {Mann}(2005)}]{2005A&A...435.1123W}
{Warmuth}, A., \& {Mann}, G. 2005, \aap, 435, 1123

\bibitem[{{Weber}(1973{\natexlab{a}})}]{1973Ap&SS..20..391W}
{Weber}, E.~J. 1973{\natexlab{a}}, \apss, 20, 391

\bibitem[{{Weber}(1973{\natexlab{b}})}]{1973Ap&SS..20..401W}
---. 1973{\natexlab{b}}, \apss, 20, 401

\bibitem[{{Wentzel}(1974)}]{1974SoPh...39..129W}
{Wentzel}, D.~G. 1974, \solphys, 39, 129

\bibitem[{{Wort}(1971)}]{1971PlPh...13..258W}
{Wort}, D.~J.~H. 1971, Plasma Physics, 13, 258

\bibitem[{{Zaqarashvili} {et~al.}(2011{\natexlab{a}}){Zaqarashvili},
  {Khodachenko}, \& {Rucker}}]{2011A&A...534A..93Z}
{Zaqarashvili}, T.~V., {Khodachenko}, M.~L., \& {Rucker}, H.~O.
  2011{\natexlab{a}}, \aap, 534, A93

\bibitem[{{Zaqarashvili} {et~al.}(2011{\natexlab{b}}){Zaqarashvili},
  {Khodachenko}, \& {Rucker}}]{2011A&A...529A..82Z}
---. 2011{\natexlab{b}}, \aap, 529, A82

\end{thebibliography}

\appendix

\section{Coefficients of the matrices $\Rm{A}_{\pm}$} \label{app:A}
	\begin{minipage}{0.5\hsize}
		\begin{equation} \label{eq:a11}
		A_{\pm,11}=\left(\omega_{\pm} \mp \Omega_{1}\right) \pm \frac{Z_{1}n_{1}\Omega_{1}}{n_{e}} +i\left(\nu_{12}+\nu_{13}\right)
		\end{equation}
	\end{minipage}
	\begin{minipage}{0.5\hsize}
		\begin{equation} \label{eq:a12}
		A_{\pm,12}=\frac{\pm Z_{2}n_{2}\Omega_{1}}{n_{e}}-i \nu_{12}
		\end{equation}
	\end{minipage}
	
	\begin{minipage}{0.5\hsize}
		\begin{equation} \label{eq:a13}
		A_{\pm,13}=\frac{\pm Z_{3}n_{3}\Omega_{1}}{n_{e}}-i \nu_{13}
		\end{equation}
	\end{minipage}
	\begin{minipage}{0.5\hsize}
		\begin{equation} \label{eq:a14}
		A_{\pm,14}=\frac{k_{x}\Omega_{1}}{e n_{e}\mu_{0}}
		\end{equation}
	\end{minipage}
	
	\begin{minipage}{0.5\hsize}
		\begin{equation} \label{eq:a21}
		A_{\pm,21}=\frac{\pm Z_{1}n_{1}\Omega_{2}}{n_{e}} -i\nu_{21}
		\end{equation}
	\end{minipage}
	\begin{minipage}{0.5\hsize}
		\begin{equation} \label{eq:a22}
		A_{\pm,22}=\left(\omega_{\pm} \mp \Omega_{2}\right)\pm \frac{Z_{2}n_{2} \Omega_{2}}{n_{e}}+i\left(\nu_{21}+\nu_{23}\right)
		\end{equation}
	\end{minipage}
	
	\begin{minipage}{0.5\hsize}
		\begin{equation} \label{eq:a23}
		A_{\pm,23}=\frac{\pm Z_{3}n_{3}\Omega_{2}}{n_{e}}-i \nu_{23}
		\end{equation}
	\end{minipage}
	\begin{minipage}{0.5\hsize}
		\begin{equation} \label{eq:a24}
		A_{\pm,24}=\frac{k_{x}\Omega_{2}}{e n_{e}\mu_{0}}
		\end{equation}
	\end{minipage}
	
	\begin{minipage}{0.5\hsize}
		\begin{equation} \label{eq:a31}
		A_{\pm,31}=\frac{\pm Z_{1}n_{1}\Omega_{3}}{n_{e}} -i\nu_{31}
		\end{equation}
	\end{minipage}
	\begin{minipage}{0.5\hsize}
		\begin{equation} \label{eq:a32}
		A_{\pm,32}=\frac{\pm Z_{2}n_{2}\Omega_{3}}{n_{e}}-i \nu_{32}
		\end{equation}
	\end{minipage}
	
	\begin{minipage}{0.5\hsize}
		\begin{equation} \label{eq:a33}
		A_{\pm,33}=\left(\omega_{\pm} \mp \Omega_{3}\right) \pm \frac{Z_{3}n_{3}\Omega_{3}}{n_{e}} + i \left(\nu_{31}+\nu_{32}\right)
		\end{equation}
	\end{minipage}
	\begin{minipage}{0.5\hsize}
		\begin{equation} \label{eq:a34}
		A_{\pm,34}=\frac{k_{x}\Omega_{3}}{e n_{e}\mu_{0}}
		\end{equation}
	\end{minipage}
	
	\begin{minipage}{0.5\hsize}
		\begin{equation} \label{eq:a41}
		A_{\pm,41}=\frac{k_{x}B_{x}Z_{1}n_{1}}{n_{e}}
		\end{equation}
	\end{minipage}
	\begin{minipage}{0.5\hsize}
		\begin{equation} \label{eq:a42}
		A_{\pm,42}=\frac{k_{x}B_{x}Z_{2}n_{2}}{n_{e}}
		\end{equation}
	\end{minipage}
	
	\begin{minipage}{0.5\hsize}
		\begin{equation} \label{eq:a43}
		A_{\pm,43}=\frac{k_{x}B_{x}Z_{3}n_{3}}{n_{e}}
		\end{equation}
	\end{minipage}
	\begin{minipage}{0.5\hsize}
		\begin{equation} \label{eq:a44}
		A_{\pm,44}=\omega \pm \frac{k_{x}^2B_{x}}{e n_{e}\mu_{0}}
		\end{equation}
	\end{minipage}

\section{Auxiliary parameters} \label{app:B}
	\begin{equation} \label{eq:Gamma}
		\Gamma=\frac{\alpha_{12}}{\rho_{1} \rho_{2}\Omega_{1} \Omega_{2}}\left(\left(\rho_{1}+\rho_{2}\right) \omega_{\pm}\mp\left(\rho_{1}\Omega_{1}+\rho_{2}\Omega_{2}\right)\right).
	\end{equation}
	
	\begin{equation*} 
		\widetilde{\Omega_{1}}=\frac{Z_{1}n_{1}\left(\Omega_{2}+\Omega_{3}\right) + Z_{2}n_{2}\left(\Omega_{1}+\Omega_{3}\right) + Z_{3}n_{3}\left(\Omega_{1}+\Omega_{2}\right)}{2 n_{e}}
	\end{equation*}
	
	\begin{equation} \label{eq:weighted_1}
		-\frac{\sqrt{\big(Z_{1}n_{1}\left(\Omega_{2}+\Omega_{3}\right) + Z_{2}n_{2}\left(\Omega_{1}+\Omega_{3}\right) + Z_{3}n_{3}\left(\Omega_{1}+\Omega_{2}\right)\big)^2-4n_{e}\left(Z_{1}n_{1}\Omega_{2}\Omega_{3} + Z_{2}n_{2}\Omega_{1}\Omega_{3} + Z_{3}n_{3}\Omega_{1}\Omega_{2}\right)}}{2 n_{e}}
	\end{equation}
	
	\begin{equation*} 
		\widetilde{\Omega_{2}}=\frac{Z_{1}n_{1}\left(\Omega_{2}+\Omega_{3}\right) + Z_{2}n_{2}\left(\Omega_{1}+\Omega_{3}\right) + Z_{3}n_{3}\left(\Omega_{1}+\Omega_{2}\right)}{2 n_{e}}
	\end{equation*}
	
	\begin{equation} \label{eq:weighted_2}
		+\frac{\sqrt{\big(Z_{1}n_{1}\left(\Omega_{2}+\Omega_{3}\right) + Z_{2}n_{2}\left(\Omega_{1}+\Omega_{3}\right) + Z_{3}n_{3}\left(\Omega_{1}+\Omega_{2}\right)\big)^2-4n_{e}\left(Z_{1}n_{1}\Omega_{2}\Omega_{3} + Z_{2}n_{2}\Omega_{1}\Omega_{3} + Z_{3}n_{3}\Omega_{1}\Omega_{2}\right)}}{2 n_{e}}
	\end{equation}

\end{document}